\documentclass[11pt]{article}
\usepackage{graphicx,amsthm,graphicx,fancyhdr,mathrsfs}
\usepackage{amsfonts}
\usepackage{amsmath}
\usepackage{amssymb}
\usepackage{url}
\usepackage[colorlinks=true,citecolor=blue]{hyperref}
\usepackage{fancyhdr}
\usepackage{indentfirst}
\usepackage{enumerate}
\usepackage{amsthm}
\usepackage{color}
\usepackage{comment}
\usepackage{dsfont}
\usepackage{natbib}
\usepackage[misc]{ifsym}
\usepackage[toc,page]{appendix}
\usepackage{multirow}
\usepackage{float}
\usepackage{bm}
\usepackage{booktabs}
\usepackage{todonotes}
\usepackage{subcaption}
\usepackage{graphicx}

\def\d{\mathrm{d}}
\newcommand{\VaR}{\mathrm{VaR}}

\newcommand{\ES}{\mathrm{ES}}

\newcommand{\X}{\mathcal{X}}

\newcommand{\E}{\mathbb{E}}
\newcommand{\R}{\mathbb{R}}
\newcommand{\N}{\mathbb{N}}

\newcommand{\M}{\mathcal{M}}

\newcommand{\p}{\mathbb{P}}

\newcommand{\q}{\mathbb{Q}}

\newcommand{\id}{\mathds{1}}

\renewcommand{\ge}{\geqslant}
\renewcommand{\le}{\leqslant}
\renewcommand{\geq}{\geqslant}
\renewcommand{\leq}{\leqslant}

\def\lawis{\buildrel \mathrm{d} \over \sim}

\newcommand{\bR}{\overline{\R}}

\newcommand{\esssup}{\mathrm{ess\mbox{-}sup}}
\newcommand{\essinf}{\mathrm{ess\mbox{-}inf}}

\theoremstyle{plain}
\newtheorem{theorem}{Theorem}

\newtheorem{proposition}{Proposition}
\theoremstyle{definition}
\newtheorem{definition}{Definition}

\newtheorem{remark}{Remark}

\DeclareMathOperator*{\argmin}{arg\,min}

\definecolor{DarkGreen}{rgb}{0.2,0.6,0.2}

\definecolor{purple}{rgb}{0.6,0.3,0.8}

\usepackage{pgfplots}
\usepackage{tikz}

\renewcommand{\cite}{\citet}

\makeatletter
\DeclareRobustCommand{\bsquare}{%
	\mathop{\vphantom{\sum}\mathpalette\bigstar@\relax}\slimits@
}

\newcommand{\bigstar@}[2]{%
	\vcenter{%
		\sbox\z@{$#1\sum$}%
		\hbox{\resizebox{.9\dimexpr\ht\z@+\dp\z@}{!}{$\m@th\dsquare$}}%
	}%
}
\makeatother

\usepackage[onehalfspacing]{setspace}

\newcommand{\dsquare}{\mathop{  \square} \displaylimits}

\begin{document}

\title{Lambda Expected Shortfall}

% \author[*]{Muqiao Huang}
% \author[**]{Qiuqi Wang}
% \author[*]{Ruodu Wang}
% \affil[*]{Department of Statistics and Actuarial Science, University of Waterloo, Canada. \texttt{m5huang@uwaterloo.ca}}
% \affil[**]{Maurice R.~Greenberg School of Risk Science, Georgia State University, U.S.A. \texttt{qwang30@gsu.edu}}

\author{Fabio Bellini\thanks{Department of Statistics and Quantitative Methods, University of Milano-Bicocca, Italy. \texttt{fabio.bellini@unimib.it}} \and Muqiao Huang\thanks{Department of Statistics and Actuarial Science, University of Waterloo, Canada. \texttt{m5huang@uwaterloo.ca}}
\and 
Qiuqi Wang\thanks{Maurice R.~Greenberg School of Risk Science, Georgia State University, USA. \texttt{qwang30@gsu.edu}}
\and Ruodu Wang\thanks{Department of Statistics and Actuarial Science, University of Waterloo, Canada.   \texttt{wang@uwaterloo.ca}}}

\maketitle

\begin{abstract}
\medskip
\noindent The Lambda Value-at-Risk ($\Lambda$-$\mathrm{VaR}$) is a generalization of the Value-at-Risk (VaR), which has been actively studied in quantitative finance. Over the past two decades, the Expected Shortfall (ES) has become one of the most important risk measures alongside VaR because of its various desirable properties in the practice of optimization, risk management,  and financial regulation. Analogously to the intimate relation between ES and VaR, we introduce the Lambda Expected Shortfall ($\Lambda$-$\mathrm{ES}$), as a generalization of ES and a counterpart to $\Lambda$-$\mathrm{VaR}$. 
Our definition of $\Lambda$-$\mathrm{ES}$
has an explicit formula and many convenient properties, and we show that it is the smallest quasi-convex and law-invariant risk measure dominating $\Lambda$-$\mathrm{VaR}$ under mild assumptions. We examine further properties of $\Lambda$-ES, its dual representation, and related optimization problems.  

~

\end{abstract}

\section{Introduction}

%In the landscape of quantitative finance and actuarial science,  efficient and robust measurement of risk is paramount. 
Financial institutions and regulators make use of sophisticated tools to quantify potential losses and manage financial exposures effectively. Among the most widely adopted risk measures are the  Value-at-Risk (VaR) and the Expected Shortfall (ES), each with distinct theoretical properties and practical implications.
VaR has long served as a standard for risk assessment due to its intuitive interpretability. However, its well-documented limitations, such as the lack of subadditivity and non-convexity for general loss distributions and inability of capturing tail risk \citep[see e.g.,][]{DEGKMRS01,MFE15,ELW18}, have spurred the development of more robust alternatives. ES, also known as the Conditional Value-at-Risk (CVaR), emerged as the most popular alternative, with desirable features such as coherence \citep{ADEH99,AT02}, convexity \citep{FS02,FR02},  optimization properties \citep{RU02, ESW22}, 
and axiomatization via portfolio concentration \citep{WZ21}, 
although it suffers from the lack of elicitability \citep{G11,Z16, KP16, FZ16}.
% A cornerstone theoretical result highlights ES's preeminence: it is the smallest law-invariant coherent risk measure dominating VaR (Delbaen, 2012). Our work further extends this understanding by demonstrating that ES is also the smallest quasi-convex and law-invariant risk measure dominating VaR, providing a crucial precedent for our generalization.

 As a flexible generalization of VaR, the class of Lambda Value-at-Risk ($\Lambda$-VaR) was introduced by \cite{FMP14}. The class of $\Lambda$-VaR offers enhanced adaptability for modeling diverse risk preferences and regulatory contexts beyond a fixed confidence level.
$\Lambda$-VaR is found to satisfy several useful properties in finance, including monotonicity, cash subadditivity, elicitability \citep{BB15}, robustness  \citep{BPR17}, and quasi-star-shapeness \citep{HWWX25}. 
 \cite{BP22} obtained an axiomatic characterization of $\Lambda$-VaR, in particular justifying the choice of $\Lambda$ to be a (weakly) decreasing function. 
 As a risk measure,
$\Lambda$-VaR has also been studied from practical aspects such as estimation and backtesting \citep{HMP18,CP18}, distributionally robust optimizations \citep{HL25}, capital allocations \citep{IPP22,L25}, and optimal insurance problems \citep{BCHW25}.
% $\Lambda$-VaR has also been studied from various other perspectives, including its robustness, elicitability, and consistency (Burzoni et al., 2017), as well as practical aspects such as estimation and backtesting (Hitaj et al., 2018; Corbetta and Peri, 2018). Further theoretical explorations have focused on its quasi-convexity, cash subadditivity, and quasi-star-shapedness (Han et al., 2025), along with robust formulations (Han and Liu, 2024), and its application in capital allocation using the Euler rule (Ince et al., 2022).
% A foundational characterization of A-VaR is provided by Bellini and Peri (2022), who formalize a crucial property called locality. This, combined with monotonicity, normalization, and weak semi-continuity, uniquely defines A-VaR for a non-increasing function $\Lambda(x)$ on $\mathbb{R}$.
While $\Lambda$-VaR successfully broadens the scope of VaR, it retains the essential drawbacks of VaR for not being convex and not being able to capture tail risk. A natural remedy for the problem is to introduce an equally flexible generalization of ES as an alternative to $\Lambda$-VaR. A suitable way of defining such a risk measure that preserves its desirable properties and strong theoretical foundations has not been found. 

This paper addresses this gap by introducing the Lambda Expected Shortfall ($\Lambda$-ES), a natural counterpart to $\Lambda$-VaR.  
There are many potential ways to generalize ES to a class of risk measures parametrized by a function $\Lambda$. 
A key consideration in ES and its generalization is its consistency with respect to portfolio diversification, modelled via convexity by \cite{FS02,FR02}. 
For general risk measures, \cite{CMMM11} argued that diversification preferences should be modelled by quasi-convexity, which is equivalent to convexity for monetary risk measures.
Keeping this property as our fundamental requirement for a generalization of ES, we find that there is one formulation that has the most advantages, inspired by a recent $\Lambda$-VaR representation result of \cite{HWWX25}. 
For a decreasing function $\Lambda$, we define $\Lambda$-ES of a random variable $X$ by
\begin{equation}\label{eq:definition}
    \sup_{x\in\R}\left(\ES_{\Lambda(x)}(X)\wedge x\right).
\end{equation}
We demonstrate that $\Lambda$-ES defined in \eqref{eq:definition} possesses several critical properties (Proposition \ref{prop:properties}), analogous to those that establish ES as an improved alternative to VaR. More importantly, 
based on a new result on the domination of ES over VaR (Theorem \ref{lem:ES-domin}),  we show that $\Lambda$-ES   is the smallest quasi-convex and law-invariant risk measure that dominates $\Lambda$-VaR (Theorem \ref{thm:rep}). This result extends the classic dominance between VaR and ES \citep{D12,FS16}.

Beyond the foundational definition and properties of $\Lambda$-ES, which are the topics of Section \ref{sec:LES}, we proceed to conduct a comprehensive analysis of this new class of risk measures. In Section \ref{sec:dual}, we obtain a dual representation of $\Lambda$-ES (Theorem \ref{thm:dual}), offering deeper insights into its theoretical structure and connections to quasi-convex cash-subadditive risk measures. In Section \ref{sec:bayes}, we explore the properties of $\Lambda$-ES  in optimization problems, both as an objective function to minimize and as a constraint to impose, and analyze various forms of convexity in relevant reformulations of ES optimization problems. 
% As standard in the risk measures literature,  the main results are formulated on the space $L^\infty$  of essentially bounded random variables.
In Section \ref{sec:L1}, results are naturally extended to the space $L^1$ of integrable random variables, sometimes under slightly stronger assumptions.
Section \ref{sec:con} concludes the paper.
% Some alternative potential formulations for $\Lambda$-ES are discussed in Appendix \ref{sec:others}, demonstrating why our proposed definition is the most robust, theoretically consistent, and desirable for risk management applications.

% The remainder of this paper is organized as follows. Section \ref{sec:review} provides a comprehensive review of VaR, $\Lambda$-VaR, and ES, laying the necessary groundwork. In particular, we show a seemingly old but novel result (Theorem \ref{lem:ES-domin}) indicating ES as the smallest quasi-convex and law-invariant risk measure dominating VaR.
% % The result is stronger than similar established results in \cite{FS16} and \cite{D12}.
% Section \ref{sec:LES} formally introduces $\Lambda$-ES, establishing its explicit representation and detailing its key theoretical properties. Section \ref{sec:dual} presents the dual representation of $\Lambda$-ES. Section \ref{sec:bayes} explores the application of $\Lambda$-ES in optimization problems. Our discussion then extends to the $L^1$ space in Section \ref{sec:L1}.
% % where most theoretical results still hold under the same or slightly stronger assumptions.
% Section \ref{sec:con} concludes the paper. Finally, Appendix \ref{sec:others} critically examines other alternative potential formulations for $\Lambda$-ES, showing they violate key properties to be desired counterparts to $\Lambda$-VaR.

\section{VaR, Lambda VaR and ES}
\label{sec:review}

\subsection{Risk measures}
\label{sec:RM-properties}

Let $L^0$ be the space of all random variables on an atomless probability space $(\Omega,\mathcal F,\p)$, $L^1$ be the space of all random variables with finite mean, and $L^\infty$ be the set of all essentially bounded random variables. 
Write $\bR=[-\infty,\infty]$ and $\R_+=[0,\infty)$. For any $n\in\N$, denote by $[n]=\{1,\dots,n\}$. For any $x,y\in\bR$, write $x\wedge y=\min\{x,y\}$, $x\vee y=\max\{x,y\}$, $x_+=x\vee 0$, and $x_-=x \wedge 0$. For any function $f:\mathbb R\to\mathbb R$ and  $x\in\mathbb R$, we write $f(x-)=\lim_{y\uparrow x}f(y)$ and $f(x+)=\lim_{y\downarrow x}f(y)$, if they exist. Let $\mathcal M_c(\R)$ denote the set of compactly supported distributions on $\R$. 
 
  A risk measure is a mapping $\rho:\X\to\bR$, where $\X$ is the space of random variables where $\rho$ is defined. Below, we list several common properties that a risk measure may satisfy. Their financial interpretation is well documented in the classsic literature (e.g., \citealp{ADEH99, FS02, FR02}). For some recent developments on these and related properties and their interpretations, see e.g., \cite{CCMTW22}, \cite{AL24}, \cite{MPST24}, and \cite{HK25}. 

The risk measure $\rho$ is called a \emph{monetary risk measure} if it satisfies
\begin{itemize}
\item[--] \emph{Monotonicity:} $\rho(X)\ge \rho(Y)$ for all $X,Y\in\X$ and $X\ge Y$ almost surely;
\item[--] \emph{Cash additivity (or translation invariance):} $\rho(X+m)=\rho(X)+m$ for all $X\in\X$ and $m\in\R$.
\end{itemize} 
 A  monetary risk measure is often required to satisfy 
\begin{itemize}
\item[--] \emph{Normalization:} $\rho(t)=t$ for all $t\in\R$.
\end{itemize}
A monetary risk measure is called \emph{coherent}  if it further satisfies\footnote{Whenever convexity or subadditivity is discussed, the range of $\rho$  includes at most one of $\infty$ and $-\infty$ to avoid $\infty-\infty$.}
\begin{itemize}
\item[--] \emph{Positive homogeneity:} $\rho(\gamma X)=\gamma\rho(X)$ for all $X\in\X$ and $\gamma\in (0,\infty)$;
\item[--] \emph{Subadditivity:} $\rho(X+Y)\le \rho(X)+\rho(Y)$ for all $X,Y\in\X$;
\end{itemize}
whereas a monetary risk measure is called a \emph{convex risk measure}  if it further satisfies
\begin{itemize}
\item[--] \emph{Convexity:} $\rho(\gamma X+(1-\gamma)Y)\le\gamma\rho(X)+(1-\gamma)\rho(Y)$ for all $X,Y\in\X$ and $\gamma\in[0,1]$.
\end{itemize}
Convexity is motivated by diversification effects in risk measurement.  
To incorporate non-constant interest rates, \cite{KR09} relaxes cash additivity to 
\begin{itemize}
\item[--] \emph{Cash subadditivity:} $\rho(X+m)\le\rho(X)+m$ for all $X\in\X$ and $m\in\R_+$.
\end{itemize} 
For cash-subadditive risk measures, \cite{CMMM11} argued that the diversification effect  is characterized by 
\begin{itemize}
\item[--] \emph{Quasi-convexity:} $\rho(\gamma X+(1-\gamma)Y)\le \max\{\rho(X),\rho(Y)\}$ for all $X,Y\in\X$ and $\gamma\in[0,1]$.
% as argued by \cite{CMMM11}, quasi-convexity is the right notion for diversification when the risk measure is not translation invariant. 
\end{itemize}
% \cite{HWWX25} further studied cash-subadditive risk measures without assuming quasi-convexity, and showed that a natural property to study with cash-subadditive risk measures is no longer star-shapeness \citep{CCMTW20} but quasi-star-shapeness, defined as follows.
Many commonly used convex risk measures (such as the Expected Shortfall defined below) also satisfy law invariance and concavity with respect to distribution mixtures. 
\begin{itemize}
    \item[--] \emph{Law invariance:} $\rho(X)=\rho(Y)$ for all $X,Y\in\X$ with the same distribution.
    \item[--] \emph{Concavity} (resp.~\emph{quasi-concavity}) \emph{in mixtures:} $\rho$ is law invariant and the function  $F\mapsto \rho(X_F)$ on $\mathcal M_c(\R)$ is concave (resp.~quasi-concave), where $X_F$ is a random variable with distribution $F\in\mathcal M_c(\R)$.
\end{itemize}
Further properties of risk measures that we will consider in this paper include
\begin{itemize}
\item[--] \emph{SSD-consistency:} $\rho(X)\ge\rho(Y)$ for all $X,Y\in\X$ and $X\succeq_{\rm icx}Y$.\footnote{Here, SSD represents second-order stochastic dominance. For $X,Y\in\X$, we say that $X$ \emph{dominates} $Y$ in \emph{increasing convex order}, denoted by $X\succeq_{\rm icx}Y$, if $\E[f(X)]\ge \E[f(Y)]$ for all increasing and convex functions $f:\R\to\R$. SSD-consistent monetary risk measures are characterized by \cite{MW20}.}
% (quasi-convexity, monotonicity, law invariance and norm-continuity together imply this).
\item[--] \emph{$L^1$-continuity:} $\rho(X_n)\to\rho(X)$ for all $X,X_1,X_2\dots\in\X\cap L^1$ and $X_n\stackrel{L^1}{\to} X$ as $n\to\infty$. 
\end{itemize}
In the definition of $L^1$-continuity, the space $\X$ may be different from $L^1$, such as $L^\infty$ or $L^0$.
%For  some background and characterizations of the properties introduced above, we refer to \citet[convexity]{FS02}, \citet[quasi-convexity]{CMMM11}, \citet[SSD-consistency]{MW20}, and \citet[concavity in mixtures]{WWW20}.

\subsection{VaR and Lambda VaR}

The \emph{Value-at-Risk} (VaR) at level $\alpha\in [0,1]$ is defined as the left quantile, namely,
$$\VaR_{\alpha}(X)=\inf \{x \in \R: \p(X\le x) \ge \alpha\},~~~X\in L^0.$$ Similarly, the \emph{upper Value-at-Risk} ($\VaR^+$) at level $\alpha\in[0,1]$ is defined as the right quantile:
$$\VaR^+_{\alpha}(X)=\inf \{x \in \R: \p(X\le x) > \alpha\},~~~X\in L^0.$$
Using these formulations, for any $X\in L^0$, $\VaR_0(X)=-\infty$, $\VaR^+_1(X)=\infty$,  $\VaR_1(X)$ is the essential supremum of $X$,
  and $\VaR^+_0(X)$ is the essential infimum of $X$.  Moreover,  $\VaR_\alpha(X),\VaR^+_\alpha(X)\in \R$ for any $\alpha \in(0,1)$.
Both versions of VaR satisfy monotonicity, cash additivity, positive homogeneity,   law invariance, and quasi-concavity in mixtures, but not quasi-convexity, concavity in mixtures, or SSD-consistency.  %Indeed, the latter three properties are equivalent for the class of distortion risk measures; see \citet[Theorem 3]{WWW20}.

Let $\Lambda:\R\to [0,1]$ be a decreasing function.  Throughout, all terms of ``increasing" and ``decreasing" are in the weak sense.   
 The  \emph{$\Lambda$-Value-at-Risk} ($\Lambda$-VaR, or {$\Lambda$-quantile}), denoted by 
$ \VaR_{\Lambda}:L^0\to \bR$, 
 is defined as 
$$
\VaR_{\Lambda}(X)= \inf \{x \in \R: \p(X\le x) \ge \Lambda(x)\}=\sup\{x \in \R: \p(X\le x) < \Lambda(x)\},~~~X\in L^0.
$$
\cite{FMP14} originally introduced $\Lambda$-VaR focusing on the case that $\Lambda$ is increasing. \cite{BB15} showed that $\Lambda$-VaR with a decreasing $\Lambda$ satisfies elicitability, and it is not true for increasing $\Lambda$ \citep{BPR17}. A more decisive result is the axiomatic justification of  
\cite{BP22} for using a decreasing function $\Lambda $. \cite{HWWX25} further showed that $\Lambda$-VaR with a decreasing $\Lambda$ is cash subadditive and hence $L^\infty$-continuous, but with an increasing $\Lambda$ even $L^\infty$-continuity fails. For these reasons, our study focuses on the case of decreasing $\Lambda$. 

VaR has two versions, and so does $\Lambda$-VaR. We define the \emph{upper $\Lambda$-VaR} ($\Lambda$-$\VaR^+$), denoted by $\VaR^+_{\Lambda}:L^0\to\bR$, as
$$
\VaR^+_{\Lambda}(X)= \inf \{x \in \R: \p(X\le x) > \Lambda(x)\}=\sup\{x \in \R: \p(X\le x) \le \Lambda(x)\},~~~X\in L^0.
$$

If $\Lambda$ is a constant $\alpha\in [0,1]$ (written $\Lambda\equiv\alpha$), then $\VaR_\Lambda=\VaR_\alpha$  and $\VaR^+_\Lambda=\VaR^+_\alpha$. 
The risk measure $\Lambda$-VaR  is monotone, but not cash additive or positively homogeneous, thus losing some usual properties of VaR.

 \citet[Theorem 1]{HWWX25} gives a representation  of $\Lambda$-VaR, which will be useful for our study. Below, we state the result and extend it to $\Lambda$-$\VaR^+$.
\begin{proposition}\label{prop:LVaR-rep}
Let $\Lambda:\R\to[0,1]$ be a decreasing function. The risk measures $\VaR_\Lambda$ and $\VaR^+_{\Lambda}$ admit the following representations:
\begin{align}
&\VaR_{\Lambda}(X)
= \sup_{x\in \R} \left(\VaR_{\Lambda(x)}(X) \wedge x\right)= \inf_{x\in \R} \left( \VaR_{\Lambda(x)}(X) \vee x \right),~~~X\in L^0,\label{eq:LVaR-rep}\\
&\VaR^+_{\Lambda}(X)
= \sup_{x\in \R} \left(\VaR^+_{\Lambda(x)}(X) \wedge x\right)= \inf_{x\in \R} \left( \VaR^+_{\Lambda(x)}(X) \vee x \right),~~~X\in L^0.\label{eq:LVaR-rep+}
\end{align}
\end{proposition}
\begin{proof}
   Equation \eqref{eq:LVaR-rep} holds directly by Theorem 1 of \cite{HWWX25} for all decreasing functions $\Lambda:\R\to[0,1]$ that are not constantly $0$. For $\Lambda\equiv 0$, we have
    $$\sup_{x\in \R} \left(-\infty \wedge x\right)= \inf_{x\in \R} \left( -\infty \vee x \right)=-\infty=\VaR_0(X).$$
    Thus \eqref{eq:LVaR-rep} holds for all decreasing functions $\Lambda:\R\to[0,1]$. To see \eqref{eq:LVaR-rep+}, using some standard relations between quantiles and distribution functions, we have $$\begin{aligned}
        \VaR^+_{\Lambda}(X) &= \sup\{x\in\R:\p(X\le x)\le \Lambda(x)\}
\\&  = \sup \{x\in \R: \VaR^+_{\Lambda(x)}(X) \ge  x\} 
= \sup_{x\in \R} \left\{ \VaR^+_{\Lambda(x)}(X) \wedge  x \right\},
    \end{aligned}$$
    and similarly,
    $$\begin{aligned}
        \VaR^+_{\Lambda}(X)  = \inf \{x\in \R: \VaR^+_{\Lambda(x)}(X) <  x\} 
=  \inf_{x\in \R} \left\{ \VaR^+_{\Lambda(x)}(X) \vee  x \right\}. 
    \end{aligned}$$
% Therefore, we have
% $$
% \VaR^+_{\Lambda}(X)   \le \sup_{x\in \R} \left\{ \VaR^+_{\Lambda(x)}(X) \wedge  x \right\} 
% \le \inf_{x\in \R} \left\{ \VaR^+_{\Lambda(x)}(X) \vee  x \right\} \le \VaR^+_{\Lambda}(X).
% $$
%This shows \eqref{eq:LVaR-rep+}. 
The proof is complete.
\end{proof}

\subsection{Expected Shortfall}
\label{sec:ES}
The standard risk measure in banking regulation, the \emph{Expected Shortfall} (ES),  can be defined via a few different formulations. 
First,  as the most standard definition, ES at level $\alpha \in [0,1]$ is defined as the mapping 
$\ES_\alpha:L^\infty\to \R$  given by
\begin{align}
\label{eq:ES1}
\ES_\alpha(X)=\frac{1}{1-\alpha} \int_\alpha^1 \VaR_\beta (X) \d \beta  \mbox{ for $\alpha\in [0,1)$}, 
\end{align}
and $\ES_1(X)=\VaR_1(X)$.
Note that $\ES_0=\E$ and the definition of $\ES_\alpha$ in \eqref{eq:ES1} can be easily extended to $L^1$. We will discuss risk measures on $L^1$ in Section \ref{sec:L1}.
An ES satisfies all properties listed in Section \ref{sec:RM-properties}.
% For any $\alpha \in [0,1]$, $\ES_\alpha$ never takes the value $-\infty$ on $L^1$. 
Second, as shown by \cite{RU02}, for $\alpha \in [0,1]$, $\ES_\alpha$ can be equivalently formulated by 
\begin{align}
\label{eq:ES2}
\ES_\alpha(X)= \min_{x\in \R}\left\{x+\frac{1}{1-\alpha}\E[(X-x)_+]\right\},~~~X\in L^\infty,
\end{align}
where we set $0/0=0$ and $x/0=\infty$ for $x>0$. The representation \eqref{eq:ES2} connects to VaR via 
\begin{align}\label{eq:VaR2}
    \argmin_{x\in \R}\left\{x+\frac{1}{1-\alpha}\E[(X-x)_+]\right\}=\left\{\begin{array}{ll}
        [\VaR_\alpha(X),\VaR^+_\alpha(X)], & \mbox{if}~\alpha\in[0,1), \\
        \VaR_1(X), & \mbox{if}~\alpha=1,
    \end{array}\right.~~~X\in L^\infty.
\end{align}
% Here, we use the convention $0/0=0$.
We call \eqref{eq:ES2} and \eqref{eq:VaR2} the Rockafellar--Uryasev (RU) formulas for VaR and ES. 
Third, it is known that the risk measure $\ES_\alpha$ is the smallest law-invariant coherent risk measure dominating $\VaR_\alpha$ \citep[][Theorem 52]{D12}.
Convexity is important and relevant for risk management, and for this reason, ES is regarded as an improvement of VaR.  
In the next result, we show that $\ES_\alpha$ is also the smallest mapping dominating $\VaR_\alpha$ satisfying quasi-convexity and law invariance.
As far as we know, this result is new, and it is based on a VaR-ES asymptotic equivalence result of \cite{WW15} and a result in \cite{EWW15} on the sum of negatively dependent sequences. Throughout the paper, we write $\rho\ge \widetilde \rho$ for mappings $\rho:\X\to\bR$ and $\widetilde\rho:\widetilde\X\to\bR$ to represent the dominance of $\rho$ over $\widetilde\rho$ on their common domain (i.e., $\rho(X)\ge\widetilde\rho(X)$ for all $X\in\X\cap\widetilde\X$), and typically we have either $\X\subseteq \widetilde \X $ or $\widetilde \X\subseteq  \X $.

\begin{theorem}
    \label{lem:ES-domin}
    The following equalities hold:
% For any $\alpha \in (0,1]$,
\begin{align}
&\ES_\alpha = \min \{\rho:L^ \infty \to \bR \mid \rho\ge \VaR_\alpha~\mbox{and $\rho$ is quasi-convex and law invariant}\},~\alpha \in (0,1].\label{eq:ES3}\\
&\ES_\alpha = \min \{\rho:L^ \infty \to \bR \mid \rho\ge \VaR^+_\alpha~\mbox{and $\rho$ is quasi-convex and law invariant}\},~\alpha \in [0,1).\label{eq:ES3-prime}
\end{align} 
% For any $\alpha \in [0,1)$,
% \begin{align}
% \label{eq:ES3-prime}
% \ES_\alpha = \min \{\rho:L^ \infty \to \bR \mid \rho\ge \VaR^+_\alpha~\mbox{and $\rho$ is quasi-convex and law invariant}\},~\alpha \in [0,1).
% \end{align} 
\end{theorem}

\begin{proof}
(i) We first prove \eqref{eq:ES3}. Let $\rho$ be quasi-convex and law invariant satisfying $\rho \ge \VaR_\alpha$.
If $\alpha=1$, it is clear that \eqref{eq:ES3} holds, because $\VaR_1$ is quasi-convex and law invariant.
Next, suppose   $\alpha \in (0,1)$. 
For any $X\in L^\infty$ with distribution $F$, we have 
\begin{align*}
 \tag*{\scriptsize \mbox{[law invariance
of $\rho$]}}  \rho(X)  
&=
\sup\left\{
\max\{\rho(X_1),\dots,\rho(X_n)\}:{X_i\lawis F,~i\in[n]}\right\}
\\
\tag*{\scriptsize \mbox{[quasi-convexity 
of $\rho$]}}~~ 
&\ge \sup\left\{\rho\left(\frac{X_1+\dots+X_n}{n}\right):{X_i\lawis F,~i\in[n]}\right\}
\\  \tag*{\scriptsize \mbox{[$\rho\ge \VaR_\alpha$]}}~~ 
&\ge  \sup\left\{\VaR_\alpha\left(\frac{X_1+\dots+X_n}{n}\right):{X_i\lawis F,~i\in[n]}\right\}
\\  \tag*{\scriptsize \mbox{[positive homogeneity of $\VaR_\alpha$]}}~~ & =
\frac 1n \sup\left\{\VaR_\alpha\left( {X_1+\dots+X_n} \right):{X_i\lawis F,~i\in[n]}\right\}
\\ \tag*{\scriptsize \mbox{[Corollary 3.7 of \cite{WW15}]}}~~&\to \ES_\alpha(X),~~~\mbox{as $n\to\infty$}.
\end{align*}
This shows that $\rho \ge \ES_\alpha$ for any $\rho$ in the set in \eqref{eq:ES3}.
Since $\ES_\alpha$ also satisfies law invariance and quasi-convexity, we know that the minimum of the set in \eqref{eq:ES3} is $\ES_\alpha$. 

(ii)   For $\alpha\in(0,1)$, the result in part (i) implies that $\ES_\alpha$ is the smallest quasi-convex and law-invariant risk measure that dominates $\VaR_\alpha$,
and since $\ES_\alpha\ge \VaR_\alpha^+\ge \VaR_\alpha$, the conclusion also holds for $\VaR_\alpha^+$.  

For $\alpha=0$ and $X\in L^\infty$ with distribution $F$, write $M=\VaR_1(X)-\VaR^+_0(X)$. For any $n\in\N$, by Corollary A.3 of \cite{EWW15}, there exist $\widetilde X_i\lawis F$, $i\in[n]$, such that
$|\sum^n_{i=1}\widetilde X_i/n-\E[X]|\le M/n.$
Hence,
$\sum^n_{i=1}\widetilde X_i/n\ge \E[X]-M/n.$
It yields that
$$\E[X]\ge \frac1n\VaR^+_0\left( {\widetilde X_1+\dots+\widetilde X_n} \right)\ge \E[X]-\frac Mn.$$
Therefore,
\begin{equation*}
    \frac{1}{n}\sup\left\{\VaR^+_0\left( {X_1+\dots+X_n} \right):{X_i\lawis F,~i\in[n]}\right\}\to\E[X],~~~\mbox{as }n\to\infty.
\end{equation*}
Hence, we have $\rho(X)\ge \E[X]$ in the same sense as the argument in part (i). As $\E$ dominates the essential infimum $\VaR^+_0$, it implies that $\E$ is the smallest quasi-convex and law-invariant risk measure that dominates $\VaR^+_0$.
\end{proof}

Theorem \ref{lem:ES-domin} is stronger than two classical results: \citet[Theorem 4.67]{FS16}, which requires $\rho$ in \eqref{eq:ES3} to be convex, monetary and Fatou-continuous,
and 
\citet[Theorem 52]{D12}, which requires $\rho$ in \eqref{eq:ES3} to be coherent.\footnote{We say a risk measure $\rho:\X\to\R$ is \emph{Fatou-continuous} if it is lower semicontinuous under bounded pointwise convergence: For all bounded $X,X_1,X_2,\dots\in\X$ such that $X_n\to X$ pointwise as $n\to\infty$, $\rho(X)\le\liminf_{n\to\infty}\rho(X_n)$.}
Both of the two results above further assumed that $\rho$ takes finite values and $\alpha \in(0,1)$, but these differences are not essential.  
We note that \eqref{eq:ES3}   fails for $\alpha=0$ because $\VaR_0=-\infty$ is quasi-convex, and the smallest quasi-convex and law-invariant risk measure dominating $\VaR_0$ is itself instead of $\ES_0=\E$. 
Similarly, 
 \eqref{eq:ES3-prime} fails for $\alpha=1$ because  $\VaR_1^+=\infty$ is quasi-convex, and 
the smallest quasi-convex and law-invariant risk measure dominating $\VaR_1^+$ is  itself  instead of $\ES_1=\VaR_1$.
% For $\alpha \in (0,1)$, the statement in Theorem \ref{lem:ES-domin} extends directly to  $L^1$, but for $\alpha=0$ and $\alpha=1$, some minor adjustments are needed, which we discuss in Section \ref{sec:L1}.

\section{Lambda ES}
\label{sec:LES}

%The main question of this paper is how to define a counterpart to $\Lambda$-VaR that is similar to ES as a counterpart to VaR. We will call this counterpart $\Lambda$-Expected Shortfall, or $\Lambda$-ES for short.

% \subsection{Basic requirements for Lambda ES}
% \label{sec:properties}

Now we turn to our main task of formulating the
$\Lambda$-Expected Shortfall ($\Lambda$-ES). 
By introducing a new class of risk measures, there should be some clear gain. Otherwise, the newly defined class is not useful. 
The following properties are also satisfied by $\Lambda$-VaR, and they will be considered as basic requirements for $\Lambda$-ES. We believe their desirability is self-evident. 
% We use a risk measure $\rho:\X\to\bR$ to show the definitions of the properties, where $\X$ is the space of random variables where $\rho$ is defined.
\begin{itemize} 
\item[--]   $\Lambda$-ES should  be parameterized only by  the function $\Lambda$.
\item[--]    $\Lambda$-ES should coincide with $\ES_\alpha$ when $\Lambda$ is equal to a constant $\alpha \in [0,1]$.
\item[--]    $\Lambda$-ES should increase  as  $\Lambda$ increases.
\item[--]    $\Lambda$-ES should be monotone and law invariant.
\end{itemize}
The next four properties are additional requirements for $\Lambda$-ES to be considered a useful alternative to $\Lambda$-VaR, and they highlight the contrasts between ES and VaR.
\begin{itemize}
\item[--]   $\Lambda$-ES should  dominate $\Lambda$-VaR. This is analogous to the dominance of ES over VaR. 
\item[--]   $\Lambda$-ES should be quasi-convex. This should be the key improvement of $\Lambda$-ES over $\Lambda$-VaR so that it captures the diversification effects.
\item[--]   $\Lambda$-ES should be SSD-consistent. This property  allows for $\Lambda$-ES to capture strong risk aversion in decision theory and to make consistent risk assessment.
% (quasi-convexity, monotonicity, law invariance and norm-continuity together imply this).

 \item[--]    $\Lambda$-ES should be 
$L^1$-continuous. 
This property models a form of robustness for law-invariant risk measures \citep{KSZ14}.
\end{itemize}
% A general definition of $\Lambda$-ES may be problematic for some decreasing $\Lambda:\R\to [0,1]$, and we can look at a subset of $\Lambda$ if needed. 

Some other properties, such as normalization, cash subadditivity,   and quasi-concavity in mixtures are also  natural from the corresponding properties of ES, although they may be less critical.
% They will all be satisfied by our proposed candidate for $\Lambda$-ES.

% Intuitively, there should be four ways, analogously to \eqref{eq:ES1}--\eqref{eq:ES4}, to define  $\Lambda$-ES.
% There is also fifth way, that will be identical to \eqref{eq:ES2} for ES but not clearly identical to \eqref{eq:ES2} 
% for $\Lambda$-ES,
%  which we will explain below.

%\subsection{The smallest quasi-convex risk measure dominating Lambda VaR}

%\subsection{A formal definition of Lambda ES}

With these desirable properties in mind, we are ready to define the $\Lambda$-ES. 
We first give the formal definition, which is inspired by the $\Lambda$-VaR representtion in Proposition \ref{prop:LVaR-rep}, and then show that it satisfies all desirable properties discussed above.  

% For a decreasing function $\Lambda:\R\to[0,1]$, a natural candidate in view of Theorem \ref{thm:rep} is 
% \begin{align}
% \label{eq:LES3}
%    \min \{\rho : L^\infty\to \bR \mid \rho\ge \VaR_\Lambda~\mbox{and $\rho$ is quasi-convex and law invariant}\}.
% \end{align} 

% Inspired by \eqref{eq:LVaR-rep}, we formally define the Lambda Expected Shortfall as follows.
\begin{definition}[$\Lambda$-Expected Shortfall]\label{def:LES}
    For a decreasing function $\Lambda:\R\to[0,1]$, the \emph{$\Lambda$-Expected Shortfall} ($\Lambda$-ES) is defined as the risk measure $\ES_\Lambda:L^\infty\to \R$ given by
    \begin{align}
    \label{eq:gate}
        \ES_{\Lambda}(X)   = \sup_{x\in \R} \left(\ES_{\Lambda(x)}(X) \wedge x\right),~~~X\in L^\infty.
    \end{align}
\end{definition}

Note the distinction in notation: $\ES_\Lambda$ is the mathematical object in \eqref{eq:gate}, whereas $\Lambda$-ES refers to the  concept, as we have been speaking of it without formal definition.

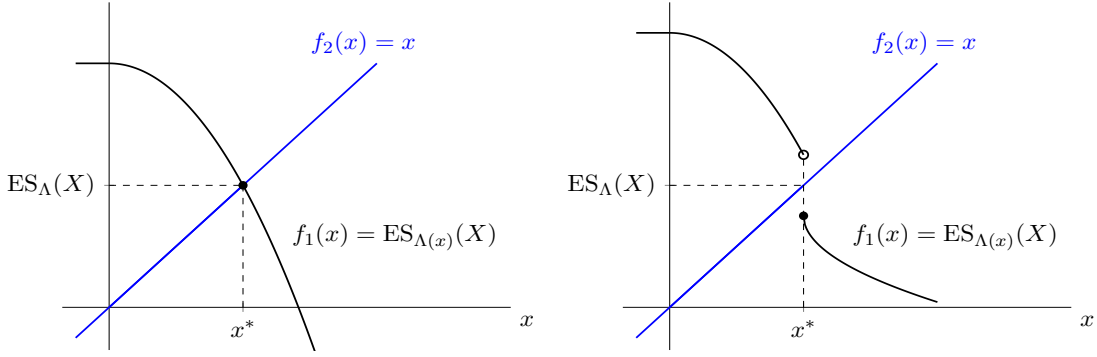
\begin{figure}[ht]
    \begin{center}
    \resizebox{0.9\textwidth}{0.28\textwidth}{
        \begin{tikzpicture}[scale=1]
            % Left Panel: Continuous Lambda
            \begin{axis}[
                title={},
                xlabel={$x$},
                ylabel={},
                xmin=-0.7, xmax=6,
                ymin=-0.7, ymax=5,
                axis lines=middle,
                axis line style={-},
                legend pos=north east,
                grid=none, % No grid
                xtick={2}, % Remove x-axis numbers
                xticklabels={$x^*$},
                ytick={2}, % Remove y-axis numbers
                yticklabels={$\ES_\Lambda(X)$},
                xlabel style={below right}, % Positioning x-axis label
                ylabel style={above left},   % Positioning y-axis label
            ]
            % ES_Lambda(x) function (decreasing and gentle non-linear)
               \addplot[thick, domain=-0.5:0, samples=100] {4};
            \addplot[thick, domain=0:4, samples=100] {4 - 0.5*(x^2)};

            % \addlegendentry{$f_1(x)=\ES_{\Lambda(x)}(X)$}
           % \node at (axis cs:0.1,4.2) [above right]{$f_1(x)=\ES_{\Lambda(x)}(X)$};
            \node at (axis cs:2.6,0.8) [above right]{$f_1(x)=\ES_{\Lambda(x)}(X)$};
            
            % f(x) = x function
            \addplot[thick, domain=-0.5:4, samples=100, blue] {x};
            \addplot[thick, domain=0:2, samples=100, blue] {x};
            % \addlegendentry{$f_2(x)=x$}
            \node[blue] at (axis cs:3.8,4) [above]{$f_2(x)=x$};

            % Intersection point
            \fill[black] (axis cs:2,2) circle (2pt);
            \draw[dashed] (axis cs:2,0) -- (axis cs:2,2);
            \draw[dashed] (axis cs:0,2) -- (axis cs:2,2);
            \end{axis}
        \end{tikzpicture}
    \hspace{0in} % Space between the two panels
        \begin{tikzpicture}[scale=1]
            % Left Panel: Continuous Lambda
            \begin{axis}[
                title={},
                xlabel={$x$},
                ylabel={},
                xmin=-0.7, xmax=6,
                ymin=-0.7, ymax=5,
                axis lines=middle,
                axis line style={-},
                legend pos=north east,
                grid=none, % No grid
                xtick={2},
                xticklabels={$x^*$},
                ytick={2},
                yticklabels={$\ES_\Lambda(X)$},
                xlabel style={below right}, % Positioning x-axis label
                ylabel style={above left},   % Positioning y-axis label
            ]
            % ES_Lambda(x) function (decreasing and gentle non-linear)
              \addplot[thick, domain=-0.5:0, samples=100] {4.5};
            \addplot[thick, domain=0:2, samples=100] {4.5 - 0.5*(x^2)}; % For x < 2
            %\addplot[dashed, domain=2:4, samples=100] {1.5-(x-2)};  % For x >= 2, decreasing
            \addplot[thick, domain=2:4, samples=100] {1.5-(x-2)^(1/2};
            % \addlegendentry{$f_1(x)=\ES_{\Lambda(x)}(X)$}
          %  \node at (axis cs:0.1,4.2) [above right]{$f_1(x)=\ES_{\Lambda(x)}(X)$};
            \node at (axis cs:2.6,0.8) [above right]{$f_1(x)=\ES_{\Lambda(x)}(X)$};
            
            % f(x) = x function
            \addplot[thick, domain=-0.5:4, samples=100, blue] {x};
            \addplot[thick, domain=0:2, samples=100, blue] {x};
            % \addlegendentry{$f_2(x)=x$}
            \node[blue] at (axis cs:3.8,4) [above]{$f_2(x)=x$};

            % Intersection point
           % \draw[thick, black] (axis cs:2,2) circle (2pt);
            \draw[thick, black] (axis cs:2,2.5) circle (2pt);
            \fill[black] (axis cs:2,1.5) circle (2pt);
            \draw[dashed] (axis cs:2,0) -- (axis cs:2,2.5);
            \draw[dashed] (axis cs:0,2) -- (axis cs:2,2);
            \end{axis}
        \end{tikzpicture}
    }

    \caption{Illustration of $\ES_\Lambda$ in Definition \ref{def:LES}; left panel shows $\ES_\Lambda$ for a continuous $\Lambda$; right panel shows $\ES_\Lambda$ for $\Lambda$ that is discontinuous at $x^*=\ES_{\Lambda}(X)$}
    \label{fig:LambdaES}
    \end{center}
\end{figure}

Figure \ref{fig:LambdaES} illustrates the definition of $\ES_\Lambda$ when the function $\Lambda$ is continuous and discontinuous, respectively. Writing $x^*=\ES_\Lambda(X)$,
we can see that    $(x^*,x^*)$ is the unique intersection point between the graph (linearly interpolated) of the function $x\mapsto \ES_{\Lambda(x)}(X)$ and the graph of the identity. From the right panel of Figure \ref{fig:LambdaES}, we can also see that whether  $x\mapsto \ES_{\Lambda(x)}(X)$ is left- or  right-continuous at $x^*$  (or neither) does not matter. 
% \vspace{-.25in}
% \textcolor{red}{add a figure here to illustrate this definition}

By definition, $\ES_\Lambda$ is finite on $L^\infty$; see also item (a) of Proposition \ref{prop:properties} below.
The next result shows that $\ES_\Lambda$  is the smallest quasi-convex and law-invariant mapping dominating $\VaR_\Lambda$.
Therefore, it is the unique formulation of the concept of $\Lambda$-ES that generalizes the ES-VaR relation in Theorem \ref{lem:ES-domin}. % Moreover, it admits 
%an alternative representation \eqref{eq:gate1}, which  can   be clearly seen  from Figure \ref{fig:LambdaES}.  
\begin{theorem}\label{thm:rep}
The following statements hold.
\begin{enumerate}[(i)]
    \item For a decreasing function $\Lambda:\R\to(0,1]$, the smallest quasi-convex and law-invariant risk measure on $L^\infty$ dominating $\VaR_\Lambda$ is $\ES_\Lambda$, that is, 
  $$
    \ES_\Lambda = \min \{\rho : L^\infty\to \bR \mid \rho\ge \VaR_\Lambda~\mbox{and $\rho$ is quasi-convex and law invariant}\}.
     $$
     \item For a decreasing function $\Lambda:\R\to[0,1)$, the smallest quasi-convex and law-invariant risk measure on $L^\infty$ dominating $\VaR^+_\Lambda$ is $\ES_\Lambda$, that is, 
  $$
    \ES_\Lambda = \min \{\rho : L^\infty\to \bR \mid \rho\ge \VaR^+_\Lambda~\mbox{and $\rho$ is quasi-convex and law invariant}\}.
     $$
\end{enumerate}
Moreover, the following identity holds for all decreasing functions $\Lambda:\R\to[0,1]$:
\begin{align}
\label{eq:gate1}
 %\sup_{x\in \R} \left(\ES_{\Lambda(x)}(X) \wedge x\right)= 
 \ES_{\Lambda}(X)  = \inf_{x\in \R} \left( \ES_{\Lambda(x)}(X) \vee x \right),~~~X\in L^\infty.
\end{align} 
\end{theorem} 
\begin{proof}
    Using \eqref{eq:LVaR-rep}--\eqref{eq:gate}, we can see that $\ES_\Lambda$
    dominates $\VaR_\Lambda$ for $\Lambda:\R\to[0,1]$ (resp.~$\VaR^+_\Lambda$ for $\Lambda:\R\to[0,1)$) because $\ES_\alpha\ge \VaR_\alpha$ for all $\alpha \in [0,1]$ (resp.~$\ES_\alpha\ge \VaR^+_\alpha$ for all $\alpha \in [0,1)$). 
    Moreover, $\ES_\Lambda$
    is law invariant by definition. Next, we show that $\ES_\Lambda$ is quasi-convex. Note that for any given $\alpha\in[0,1]$, $\ES_\alpha$ is quasi-convex. Further, an increasing transform of a quasi-convex function is quasi-convex, as well as the supremum of a set of quasi-convex functions. Using these facts,
    \begin{align*}
    \ES_{\Lambda(x)} \mbox{ is quasi-convex for each $x\in \R$}&\implies \ES_{\Lambda(x)}\wedge x \mbox{ is quasi-convex for each $x\in \R$}
    \\&\implies \sup_{x\in \R} \left(\ES_{\Lambda(x)}\wedge x\right) \mbox{ is quasi-convex.}
\end{align*}
Therefore, $\ES_\Lambda$ is a quasi-convex and law-invariant risk measure dominating $\VaR_{\Lambda}$ for $\Lambda:\R\to[0,1]$ 
% \com{$(0,1]$?}
(resp.~$\VaR^+_{\Lambda}$ for $\Lambda:\R\to[0,1)$).

Next, for $\Lambda:\R\to(0,1]$, we show that for any $\rho$ that is quasi-convex, law invariant, and satisfying $\rho\ge\VaR_{\Lambda}$, 
it must be $\rho\ge \ES_{\Lambda}$. 
For any $X \in L^\infty$, we have:
\begin{align*}
    \rho(X) \geq \VaR_\Lambda(X)
    &\implies \rho(X) \geq \sup_{x\in \R}  \left(\VaR_{\Lambda(x)}(X)\wedge x \right) \\
    % &\implies \mbox{for all $x\in \R:~$} \rho(X) \geq \left(\VaR_{\Lambda(x)}(X)\wedge x \right)  \\
    &\implies \mbox{for all $x\in \R:~$} \rho(X) \geq \VaR_{\Lambda(x)}(X) \mbox{ or } \rho(X) \geq x   \\
   \mbox{\scriptsize [Theorem \ref{lem:ES-domin}]} &\implies \mbox{for all $x\in \R:~$} \rho(X) \geq \ES_{\Lambda(x)}(X) \mbox{ or } \rho(X) \geq x \\
    &\implies \rho(X) \geq \sup_{x\in \R} \left(\ES_{\Lambda(x)}(X)\wedge x \right) = \ES_\Lambda(X). 
\end{align*}
For $\Lambda:\R\to[0,1)$, for any $\rho$ that is quasi-convex, law invariant, and satisfying $\rho\ge\VaR^+_{\Lambda}$, we have $\rho(X)\ge\ES_\Lambda(X)$ for any $X\in L^\infty$ with the same argument as above by replacing $\VaR_{\Lambda(x)}$ by $\VaR^+_{\Lambda(x)}$ for all $x\in\R$.
This completes the proof of statements (i) and (ii).
The final statement in \eqref{eq:gate1} follows from \eqref{eq:LVaR-rep}, by noting that an ES curve $\alpha\mapsto \ES_\alpha(X)$ for   $X \in L^\infty$ can be written as a VaR (resp.~$\VaR^+$) curve $\alpha\mapsto \VaR_\alpha(Y)$ for some   $Y\in L^0$ on $\alpha\in(0,1]$ (resp.~$\alpha\in[0,1)$); see e.g., Lemma 4.5 of \cite{BMW22}. 
\end{proof}

%\begin{remark}\label{rem:simple}
In Theorem \ref{thm:rep},
the reason to exclude $0$ (resp.~$1$) in part (i) (resp.~part (ii)) from the range of $\Lambda $ is the same as that in Theorem \ref{lem:ES-domin}, where $0$ (resp.~$1$) is excluded from the domination of $\ES_\alpha$ over $\VaR_\alpha$ (resp.~$\VaR_\alpha^+$), explained at the end of Section \ref{sec:ES}.

An immediate consequence of \eqref{eq:gate}--\eqref{eq:gate1} is that, for any $X\in L^\infty $ and $x\in \R$, we have
\begin{align}\label{eq:general}
    \ES_{\Lambda(x+)}(X)\le x\le \ES_{\Lambda(x-)}(X)~ \iff ~\ES_{\Lambda}(X)=x.
\end{align}
This is also illustrated in Figure \ref{fig:LambdaES}. 
As a result of \eqref{eq:general}, for any $X\in L^\infty $ and $x\in \R$, we have
\begin{align}
\label{eq:simple}
 \ES_{\Lambda(x)}(X)=x~ \implies ~\ES_{\Lambda}(X)=x;
\end{align}
moreover, if $\Lambda$ is continuous, then 
\eqref{eq:simple} becomes an equivalence.
These relations   will be convenient in some proof arguments. 
%\end{remark}

\begin{remark}
    As standard in the risk measures literature,  the main results are formulated on the space $L^\infty$ of essentially bounded random variables. 
    The results in
    Theorems \ref{lem:ES-domin} and \ref{thm:rep} 
    hold on the space $L^1$ of integrable random variables for $\alpha $ and $\Lambda$ taking values in $(0,1)$, following the same proof arguments, but for $\alpha=0$ and $\alpha=1$, some minor adjustments are needed. 
    It is straightforward to see that 
    all other results  hold on $L^1$. We discuss the details in Section \ref{sec:L1}.
\end{remark}

\begin{remark}\label{rem:cont}
    Let $\Lambda:\R\to[0,1]$ be a decreasing function and $\rho_\Lambda=\VaR_\Lambda$, $\VaR^+_\Lambda$ or $\ES_\Lambda$. The supremum in
    $\rho=\sup_{x\in\R}\{\rho_{\Lambda(x)}\wedge x \}$
    is a maximum when $\Lambda$ is left-continuous; similarly the infimum in
    $\rho=\inf_{x\in\R}\{\rho_{\Lambda(x)}\vee x \}$
     is a minimum when $\Lambda$ is right-continuous; see Figure \ref{fig:LambdaES} for an illustration. %These arguments apply to  the representations in \eqref{eq:LVaR-rep}--\eqref{eq:gate1}.
\end{remark}
   
It is clear that $\ES_\Lambda$ is parameterized only by the function $\Lambda$, and $\ES_\Lambda=\ES_\alpha$ when $\Lambda\equiv\alpha$ for some $\alpha\in(0,1)$. It satisfies all other desirable properties as a good candidate for $\Lambda$-$\ES$  as discussed in the beginning of the section, which we summarize in the following result.

\begin{proposition}\label{prop:properties}
    For any decreasing functions $\Lambda,\Lambda':\R\to[0,1]$, the risk measure $\ES_\Lambda$ satisfies the following properties:
(a) $\ES_\Lambda\ge \ES_{\Lambda'}$ when $\Lambda\ge \Lambda'$;
(b) $\ES_\Lambda$ is monotone;
(c) $\ES_\Lambda\ge \VaR_\Lambda$;
(d)  $\ES_\Lambda$ is quasi-convex; 
(e)  $\ES_\Lambda$ is normalized; 
(f)  $\ES_\Lambda$ is cash subadditive;
(g)  $\ES_\Lambda$ is SSD-consistent;
(h) $\ES_\Lambda$ is quasi-concave in mixtures;
(i) $\ES_\Lambda$ is $L^1$-continuous when $\Lambda$ takes values in $[0,1)$.
\end{proposition}
\begin{proof}
    Items (a) and (b) are straightforward because $\ES_\alpha(X)$ is monotone (increasing) in both $\alpha$ and $X$, and the supremum of monotone transformations of $\ES_\alpha(X)$ is also monotone.
    Items (c) and (d) are implied by Theorem \ref{thm:rep}. Item (e) follows from \eqref{eq:simple}.

    To see item (f), for $c\in\R_+$ and $X\in L^\infty$, we have
    % \com{I see the 3rd line as using $a\wedge b = ((a-c)\wedge b-c))+c$, a change of variable not needed.}
    \begin{align*}
        \ES_\Lambda(X+c)=\sup_{x\in\R}\{\ES_{\Lambda(x)}(X+c)\wedge x\}
        &=\sup_{x\in\R}\{(\ES_{\Lambda(x)}(X)+c)\wedge x\} \tag*{\scriptsize \mbox{[cash additivity of ES]}}\\
        % &=\sup_{x\in\R}\{\ES_{\Lambda(x)}(X)\wedge (x-c)\}+c\\
        &\le \sup_{x\in\R}\{\ES_{\Lambda(x)}(X)\wedge x\}+c=\ES_\Lambda(X)+c.
    \end{align*}

    Item (g) follows by applying  Lemma 4 of 
    \cite{HWWX25}, using the fact that $\ES_\Lambda$ is cash subadditive, monotone, quasi-convex, and law invariant. Cash subadditivity is proved in item (e). Law invariance of $\ES_{\Lambda}$ is clear from the representation in \eqref{eq:gate} and the law invariance of ES.

For item (h), we  first note that  $\ES_\alpha$ is concave in mixtures  \citep[Theorem 3]{WWW20} for each $\alpha \in [0,1]$. 
Since quasi-concavity is preserved under increasing transforms, we know that $\ES_\alpha \vee x$ is also quasi-concave in mixtures. 
By using \eqref{eq:gate1} and the fact that the infimum of quasi-concave functions is quasi-concave,  we know that $\ES_\Lambda$ is quasi-concave in mixtures.

    To prove item (i), 
    first note that  $\ES_\alpha$ is $L^1$-continuous \citep[e.g.,][Corollary 7.10]{R13} for each $\alpha \in [0,1)$. Take any random variable $X$ and any sequence $(X_n)_{n\in \N}$ in $L^\infty$ such that $X_n\to X$ in $L^1$ as $n\to\infty$.
    Let $f_n:x \mapsto \ES_{\Lambda(x)}(X_n)-x$
    and $f:x \mapsto \ES_{\Lambda(x)}(X)-x$.
    By \eqref{eq:gate}, for any $y,z$ with 
   $y<\ES_{\Lambda}(X)<z$, we have $f(y)>0 >f(z)$. Therefore,  because $f_n\to f$ pointwise,
we have  $f_n(y)>0>f_n(z)$    for $n$ large enough. This implies $y\le\ES_{\Lambda}(X_n)\le z$ via \eqref{eq:general}. Since $y,z$ are arbitrarily close to  $\ES_{\Lambda}(X)$, we know  
    $\ES_{\Lambda}(X_n)
    \to  \ES_{\Lambda}(X)$. 
\end{proof}

By item (a) of Proposition \ref{prop:properties}, it is straightforward that $\ES_{\Lambda}$ is bounded above by $\ES_1$ and below by $\E$ on $L^\infty$, which also holds when $\ES_{\Lambda}$  is formulated on larger spaces such as $L^1$.
The assumption that $\Lambda$ does not take the value $1$ in item (i) is not dispensable, noting that $\ES_1$ is not $L^1$-continuous. 

The next result shows that although $\Lambda$-ES is quasi-concave in mixtures and quasi-convex, it is neither concave in mixtures nor  convex in general, unless it is an ES. This result  also highlights the fact that quasi-convexity and convexity are different in strength for cash-subadditive risk measures, although they coincide for monetary risk measures, as shown by \cite{CMMM11}.

\begin{proposition}\label{prop:convex}
    For any decreasing function $\Lambda:\R\to[0,1]$, the following are equivalent.
    \begin{enumerate}[(i)]
        \item The risk measure $\ES_{\Lambda}$ is convex.
        \item The risk measure $\ES_{\Lambda}$ is concave in mixtures.
        \item The function $\Lambda$ is constant on $\R$.
    \end{enumerate} 
\end{proposition}
\begin{proof}
    ``(iii) $\Rightarrow$ (i)" and ``(iii) $\Rightarrow$ (ii)" follow from the facts that ES is convex and ES is concave in mixtures.
    To prove ``(i) $\Rightarrow$ (iii)", suppose that $\ES_\Lambda$ is convex and for contradiction that $\Lambda$ is not constant on $\R$. There exist $x>y$ with $\Lambda(x-)<\Lambda((x+y)/2)\le \Lambda(y)$. Take $X,Y\in L^\infty$ with $1-\p(X=x)=\p(X=y)=\Lambda((x+y)/2)$ and $Y=y$. It follows that $\ES_\Lambda(Y)=y$ and $\ES_{\Lambda}((X+Y)/2)=(x+y)/2$.
    Because $\Lambda(x-)<\Lambda((x+y)/2)$, we have $\ES_{\Lambda(x-)}(X)<x$. By \eqref{eq:general}, we have $\ES_\Lambda(X)<x$.
    % \begin{align*}
    %     &\ES_\Lambda(X)\\
    %     &=\sup_{z\in\R}\{\ES_{\Lambda(z)}(X)\wedge z\}\\
    %     &=\sup\left\{\frac{\left(\Lambda\left(\frac{x+y}{2}\right)-\Lambda(z)\right)y+\left(1-\Lambda\left(\frac{x+y}{2}\right)\right)x}{1-\Lambda(z)}\wedge z:\Lambda(z)<\Lambda\left(\frac{x+y}{2}\right),~z\in\R\right\}\tag*{\mbox{\scriptsize $\left[\Lambda(x)<\Lambda\left(\frac{x+y}{2}\right)\right]$}}\\
    %     &=\frac{\left(\Lambda\left(\frac{x+y}{2}\right)-\Lambda(x^*-)\right)y+\left(1-\Lambda\left(\frac{x+y}{2}\right)\right)x}{1-\Lambda(x^*-)}\wedge x^*<x. \tag*{\mbox{\scriptsize [$x^*=\ES_\Lambda(X)$, $\Lambda(x^*-)<\Lambda\left(\frac{x+y}{2}\right)$]}}
    % \end{align*}
    It follows that $\ES_\Lambda(X)/2+\ES_\Lambda(Y)/2<\ES_{\Lambda}((X+Y)/2)$, contradicting the convexity of $\ES_\Lambda$.
    Therefore, $\Lambda$ is constant on $\R$.

   ``(ii) $\Rightarrow$ (iii)": Suppose that $\Lambda$ is not constant on $\R$. Since 
  $\Lambda$ is bounded, it cannot be concave. 
  Hence, there exist distinct points $x,y,z\in \R$ and $\theta\in (0,1)$
  such that $z= \theta x+ (1-\theta) y $ and $\Lambda (z) < \theta  \Lambda(x) 
  + (1-\theta) \Lambda(y).
  $ By the continuity of linear functions, there exists $\gamma\in (0,1)$ in any neighborhood of $\theta$
  such that $z< \gamma x+ (1-\gamma) y $ and $\Lambda (z) < \gamma  \Lambda(x) 
  + (1-\gamma) \Lambda(y).
  $  Write  $p=\Lambda(x)$,  $q = \Lambda (y)$ and $r = \gamma p + (1-\gamma) q$. 
Take independent events $A,B,C\in \mathcal F$ such that $\p(A)=1-p$, $\p(B)=1-q$, and $\p(C)=\gamma$.   
For some constant $K > \max\{-x,-y\}$ (to be determined later), let
    $$X = x\mathbf{1}_{A}-K\mathbf{1}_{A^c},~~ Y = y\mathbf{1}_{B}-K\mathbf{1}_{B^c}, \mbox{~~and~~} Z = \mathbf{1}_C X + \mathbf{1}_{C^c}Y.$$
    We can calculate $\ES_{p}(X) = x$ and $\ES_{q}(Y) = y$. By \eqref{eq:simple}, we have $\ES_\Lambda(X) = x$ and $\ES_\Lambda(Y) = y$. Note that the distribution of $Z$ is the mixture of those of $X$ and $Y$ with weights $\gamma$ and $(1-\gamma)$ respectively. We will show $\ES_\Lambda(Z) \le  z$ for large $K$, which, together with $z <\gamma \ES_\Lambda(X) +(1-\gamma) \ES_\Lambda(Y)  $,  disproves the concavity in mixtures of $\ES_{\Lambda}$. 

    Because $\p(Z=-K)= \gamma \p(X=-K) + (1-\gamma) \p(Y=-K) = \gamma p+(1-\gamma)q =  r$ and  $\Lambda(z) < r$, 
    $$
  \ES_{\Lambda(z)} (Z) = 
   \frac{1}{1-\Lambda(z)} \left(-(r-\Lambda(z))K+    {\int_{r}^1 \VaR_\beta (Z)\,\d \beta} \right)\le    -K \frac{r-\Lambda(z)}{1-\Lambda(z)} +    \frac{1-r}{1-\Lambda(z)}  \max\{x,y\} ,
    $$
    which tends to $-\infty$ as $K\to\infty$.
    In particular, for some $K$ large enough, we have $\ES_{\Lambda(z)}(Z) \le z$.  Using \eqref{eq:gate1}, we get $\ES_\Lambda(Z) \le z$. 
\end{proof}

\begin{remark}
Another feature of ES and VaR is that they are  tail risk measures in the sense of \cite{LW21}.
More precisely, for $\alpha \in (0,1)$, 
an $\alpha$-tail risk measure is a risk measure $\rho$  such that $\rho(X)=\rho(Y)$ when the left quantile functions of $X$ and $Y$  coincide on $(\alpha,1)$.
It is straightforward to verify that $\ES_{\Lambda}$ (resp.~$\VaR_{\Lambda}$) is an $\alpha$-tail risk measure if and only if $\Lambda\ge \alpha$ (resp.~$\Lambda> \alpha$) on $\R$.
\end{remark}

\section{Dual representation}
\label{sec:dual}

We now study the dual representation of $\ES_\Lambda$ as a quasi-convex and cash-subadditive risk measure, in the form of \cite{CMMM11}. 
Denote by $\M_{1,f}=\M_{1,f}(\Omega,\mathcal F,\mathbb{P})$ the set of all finitely additive probability measures that are absolutely continuous with respect to $\p$. The following result shows the dual representation of $\ES_\Lambda$ as a direct consequence of its definition in \eqref{eq:gate}.  

\begin{theorem}\label{thm:dual}
     For any decreasing function $\Lambda:\R\to[0,1]$, the risk measure $\ES_\Lambda$ adopts the following representation:
     \begin{equation}\label{eq:dual}
         \ES_\Lambda(X)=\sup_{\q\in\M_{1,f}} R(\E_\q [X],\q), ~~~X\in L^\infty,
     \end{equation}
     where for $(t,\q)\in \R\times \mathcal M_{1,f}$,
     \begin{align}\label{eq:R}
        R(t,\q)=\sup_{x\in\R}\left\{t\wedge x~:~\Lambda(x)\ge 1-\frac{\mathrm{d}\p}{\mathrm{d}\q},~\q\mbox{-almost surely}\right\},
     \end{align}
     where we write $\d\p/\d\q=1/(\d\q/\d\p)$ with $1/0=\infty$. 
     Moreover, the following statements hold. 
     \begin{enumerate}[(i)]
        \item The supremum in \eqref{eq:dual}  is  a maximum if $\Lambda$ is left-continuous.
         \item $(t,\q)\mapsto R(t,\q)$ is upper semicontinuous, quasi-concave, and increasing in $t$;
         \item $\inf_{t\in\R} R(t,\q)=\inf_{t\in\R} R(t,\q^\prime)$ for all $\q,\q^\prime\in \M_{1,f}$;
         \item $R(t_1,\q)-R(t_2,\q)\le t_1-t_2$ for all $t_1\ge t_2$ and $\q\in\M_{1,f}$.
     \end{enumerate}
     % Moreover,
     % \begin{equation}\label{eq:CMMM11}
     %     R(t,\q)=\inf\{\ES_\Lambda(Y):\E_\q [Y]=t\},~(t,\q)\in \R\times \mathcal M_{1,f}.
     % \end{equation}
     % \begin{equation}\label{eq:dual}
     % \ES_\Lambda(X)=\inf_{x\in\R}\{\sup_{\q\in \mathcal P_{\Lambda(x)}}\E_\q[X]\vee x\}=\inf_{x\in\R}\sup_{\q\in \mathcal P_{\Lambda(x)}}\{\E_\q[X]\vee x\},~~X\in L^1,
     % \end{equation}
     % where 
     % \begin{equation*}
     % % \label{eq:prob_set}
     %     \mathcal P_{\Lambda(x)}=\left\{\q\in\M_{1,f}~:~\frac{\mathrm{d} \q}{\mathrm{d} \p}\le \frac{1}{1-\Lambda(x)}~\p\mbox{-almost surely}\right\},~~x\in\R.
     % \end{equation*}
\end{theorem}
\begin{proof}
    % \begin{align*}
    %     \ES_\Lambda(X)=\inf_{x\in\R}(\ES_{\Lambda(x)}(X)\vee x)
    %     =\inf_{x\in\R}\left\{\sup_{\q\in\mathcal P_{\Lambda(x)}}\E_{\q}[X]\vee x\right\}
    %     =\inf_{x\in\R}\sup_{\q\in\mathcal P_{\Lambda(x)}}\left\{\E_{\q}[X]\vee x\right\},
    % \end{align*}
    % where the second equality is due to Theorem 4.52 of \cite{FS16}.
    % We have
    % \begin{align*}
    %     \min_{x\in\R}\sup_{\q\in\mathcal P_{\Lambda(x)}}\left\{\E_{\q}[X]\vee x\right\}
    %     &=\min_{x\in\R}\sup_{\q\in\mathcal P}\left\{\E_{\q}[X]\vee x~:~Q\ll P,~\frac{\mathrm{d} \q}{\mathrm{d} \p}\le \frac{1}{1-\Lambda(x)}~\p\mbox{-almost surely}\right\}\\
    %     &\ge\sup_{\q\in\mathcal P}\min_{x\in\R}\left\{\E_{\q}[X]\vee x~:~Q\ll P,~\frac{\mathrm{d} \q}{\mathrm{d} \p}\le \frac{1}{1-\Lambda(x)}~\p\mbox{-almost surely}\right\}\\
    %     &=
    % \end{align*}
    Define 
    \begin{equation*}
     % \label{eq:prob_set}
         \mathcal P_{\Lambda(x)}=\left\{\q\in\M_{1,f}~:~\frac{\mathrm{d} \q}{\mathrm{d} \p}\le \frac{1}{1-\Lambda(x)},~\p\mbox{-almost surely}\right\},~~x\in\R.
     \end{equation*}
    For any $X\in L^1$ and $x\in\R$, we have
    % \begin{align}
    %     &\sup_{\mathbb{Q}\in\mathcal P_{\Lambda(x)}} \E_\mathbb{Q}[X]\wedge x\nonumber\\
    %     &=\sup\left\{\E_\mathbb{Q}[X]:\mathbb{Q}\in\mathcal P_{\Lambda(x)},\E_\mathbb{Q}[X]\ge x\right\}\vee \sup\left\{\E_\mathbb{Q}[X]:\mathbb{Q}\in\mathcal P_{\Lambda(x)},\E_\mathbb{Q}[X]< x\right\} \wedge x\nonumber\\
    %     &=\sup\left\{\E_\mathbb{Q}[X]\wedge x:\mathbb{Q}\in\mathcal P_{\Lambda(x)},\E_\mathbb{Q}[X]\ge x\right\}\vee \sup\left\{\E_\mathbb{Q}[X]\wedge x:\mathbb{Q}\in\mathcal P_{\Lambda(x)},\E_\mathbb{Q}[X]< x\right\}\nonumber\\
    %     &=\sup_{\mathbb{Q}\in\mathcal P_{\Lambda(x)}}\left\{\E_{\mathbb{Q}}[X]\wedge x\right\}\label{eq:min}
    % \end{align}
    % Then we have
    \begin{align*}
        \ES_\Lambda (X)&=\sup_{x\in\R}\left (\ES_{\Lambda(x)}(X)\wedge x\right )\\
        &=\sup_{x\in\R}\left\{\max_{\q\in\mathcal P_{\Lambda(x)}}\E_{\q}[X]\wedge x\right\}\tag*{\scriptsize \mbox{[Theorem 4.52 of \cite{FS16}]}}\\
        &=\sup_{x\in\R}\max_{\q\in\mathcal P_{\Lambda(x)}}\left\{\E_{\q}[X]\wedge x\right\}\\
        &=\sup_{\q\in\M_{1,f}}\sup_{x\in\R}\left\{\E_\q[X]\wedge x~:~\Lambda(x)\ge 1-\frac{1}{\mathrm{d}\q/\d\p},~\p\mbox{-almost surely}\right\},
    \end{align*}
    where the supremum over $x\in\R$ can be changed to a maximum when $\Lambda$ is left-continuous. This implies statement (i).
    Further because
    \begin{align*}
        \q\left(\Lambda(x)\ge 1-\frac{\d\p}{\d\q}\right)=1&\iff\E_\p\left[\mathbf{1}_{\left\{\Lambda(x)\ge 1-\frac{1}{\d\q/\d\p}\right\}}\frac{\d\q}{\d\p}\right]=1\\
        &\iff \p\left(\left\{\frac{\d\q}{\d\p}=0~\mbox{or}~\Lambda(x)\ge 1-\frac{1}{\d\q/\d\p} \right\}\right)=1\\
        &\iff\p\left(\Lambda(x)\ge 1-\frac{1}{\d\q/\d\p}\right)=1,
    \end{align*}
    we have
    $$\ES_\Lambda (X)=\sup_{\q\in\M_{1,f}}\sup_{x\in\R}\left\{\E_\q[X]\wedge x~:~\Lambda(x)\ge 1-\frac{\d \p}{\mathrm{d}\q},~\q\mbox{-almost surely}\right\}.$$
    
    Now it remains to check statements (ii)-(iv).
    
    (ii) Upper semicontinuity can be seen by showing that for all $t_0\in\R$,
    $$\begin{aligned}
        \limsup_{t\to t_0} R(t,\q)&= \lim_{\delta\downarrow 0}\sup_{x\in\R}\left\{(t_0+\delta)\wedge x~:~\Lambda(x)\ge 1-\frac{\mathrm{d}\p}{\mathrm{d}\q},~\q\mbox{-almost surely}\right\}\\
        &=\lim_{\delta\downarrow 0}\sup_{x\in\R}\left\{t_0\wedge x~:~\Lambda(x)\ge 1-\frac{\mathrm{d}\p}{\mathrm{d}\q},~\q\mbox{-almost surely}\right\}=R(t_0,\q).
    \end{aligned}$$
    Monotonicity is straightforward, and quasi-concavity is implied by monotonicity.
    Statement (iii) is clear because $\inf_{t\in\R}R(t,\q)=-\infty$ for all $\q\in\M_{1,f}$.
    
    (iv) For all $t_1\ge t_2$ and $\q\in\M_{1,f}$,
    $$\begin{aligned}
        R(t_1,\q)-R(t_2,\q)=&~t_1\wedge \sup_{x\in\R}\left\{ x~:~\Lambda(x)\ge 1-\frac{\mathrm{d}\p}{\mathrm{d}\q},~\q\mbox{-almost surely}\right\}\\
        &-t_2\wedge \sup_{x\in\R}\left\{ x~:~\Lambda(x)\ge 1-\frac{\mathrm{d}\p}{\mathrm{d}\q},~\q\mbox{-almost surely}\right\}\le t_1-t_2.
    \end{aligned}$$
    The proof is complete. 
\end{proof}

\begin{remark}\label{rem:CMMM11}
    Suppose that $\Lambda:\R\to[0,1]$ is decreasing and left-continuous. The function $R$ we obtained in \eqref{eq:R} is a special case of that obtained by Theorem 3.1 of \cite{CMMM11} for quasi-convex cash-subadditive risk measures:
    \begin{equation}\label{eq:CMMM11}
         R(t,\q)=\inf\{\ES_\Lambda(Y):\E_\q [Y]=t\},~(t,\q)\in \R\times \mathcal M_{1,f}.
     \end{equation}
    Theorem \ref{thm:dual} automatically implies \eqref{eq:CMMM11}. Below, we show another self-contained proof for \eqref{eq:CMMM11} to provide more mathematical insight. This proof   can be seen as an alternative proof for Theorem \ref{thm:dual} with $\Lambda$ being left-continuous. 
    For any $Y\in L^\infty$, due to boundedness of $Y$, there exist $a,b\in\R$ with $a<b$, such that     \begin{equation}\label{eq:compact}
        \ES_\Lambda(Y)=\sup_{y\in [a,b]}\left(\ES_{\Lambda(y)}(Y)\wedge y\right).
    \end{equation}
    For any $y_1,y_2\in[a,b]$ with $y_1\le y_2$ and $\gamma\in[0,1]$, because $y\mapsto \ES_{\Lambda(y)}$ is decreasing, we have
    $$\begin{aligned}
        &\ES_{\Lambda(\gamma y_1+(1-\gamma)y_2)}(Y)\wedge \left(\gamma y_1+(1-\gamma)y_2\right)\\
        &=\left\{\begin{array}{ll}
        \ES_{\Lambda(\gamma y_1+(1-\gamma)y_2)}(Y)\ge \ES_{\Lambda(y_2)}(Y), & \mbox{if}~~ \ES_{\Lambda(\gamma y_1+(1-\gamma)y_2)}(Y)\le \gamma y_1+(1-\gamma)y_2, \\
        \gamma y_1+(1-\gamma)y_2 \ge y_1, & \mbox{otherwise}
        \end{array}\right.\\
        &\ge \left(\ES_{\Lambda(y_1)}(Y)\wedge y_1\right) \wedge \left(\ES_{\Lambda(y_2)}(Y)\wedge y_2\right).
    \end{aligned}$$
    Thus the function $y\mapsto \ES_{\Lambda(y)}(Y)\wedge y$ is quasi-concave. For any $(t,\q)\in \R\times \mathcal M_{1,f}$, it is clear that the set $\{Y\in L^\infty: \E_\q [Y]=t\}$ is convex and the mapping $Y\mapsto \ES_{\Lambda(y)}(Y)\wedge y$ is convex due to convexity of $\ES$. Hence,
    \begin{align*}
        &\inf\left\{\ES_\Lambda(Y):\E_\q [Y]=t\right\}\\
        &=\inf\left\{\sup_{y\in [a,b]}\left(\ES_{\Lambda(y)}(Y)\wedge y\right):\E_\q [Y]=t\right\}\tag*{\scriptsize \mbox{[by \eqref{eq:compact}]}}\\
        % &=\sup_{y\in[a,b]}\inf_{\E_\q[Y]=t}\left(\ES_{\Lambda(y)}(Y)\wedge y\right)\tag*{\scriptsize \mbox{[minimax theorem \citep{F53}]}}\\
        &=\sup_{y\in[a,b]}\left(\inf_{\E_\q[Y]=t}\ES_{\Lambda(y)}(Y)\wedge y\right)\tag*{\scriptsize \mbox{[minimax theorem \citep{F53}]}}\\
        &=\sup_{y\in[a,b]}\left(\inf_{c\in\R}\inf_{Y\in L^\infty}\left(\ES_{\Lambda(y)}(Y)-c(\E_\q[Y]-t)\right)\wedge y\right)\tag*{\scriptsize \mbox{[Lagrange duality]}}\\
        &=\sup_{y\in[a,b]}\left(\inf_{c\in\R}\left(ct-\alpha(c\q)\right)\wedge y\right)\tag*{\scriptsize \mbox{$\left[\displaystyle \alpha(\q)=\sup_{Y\in L^\infty}(\E_\q[Y]-\ES_{\Lambda(y)}(Y))\right]$}}\\
        &=\sup_{y\in[a,b]}\left(\left(t-\alpha(\q)\right)\wedge y\right)\tag*{\scriptsize \mbox{[cash additivity of $\ES$]}}.
    \end{align*}
    For all $\q\in\M_{1,f}$ and $y\in[a,b]$, by Corollary 4.19 and Theorem 4.52 of \cite{FS16}, we have $\alpha(\q)=0$ if $\d\q/\d\p\le 1/(1-\Lambda(y))$, $\p$-almost surely and $\alpha(\q)=\infty$ otherwise. Therefore, we have
    $$\sup_{y\in[a,b]}\left(\left(t-\alpha(\q)\right)\wedge y\right)=\max_{x\in\R}\left\{t\wedge x~:~\Lambda(x)\ge 1-\frac{\mathrm{d}\p}{\mathrm{d}\q},~\q\mbox{-almost surely}\right\}.$$
\end{remark}

\section{The Rockafellar--Uryasev formula and optimization}
\label{sec:bayes}

\subsection{Representing Lambda ES as a minimization}
\label{sec:5RU}

The well-known relation between VaR and ES obtained by 
\cite{RU02} as shown in \eqref{eq:ES2} provides a promising solution to ES-based optimization problems.
Let $I(\bR)$ be the set of all closed real intervals, including intervals of $[\ell,\infty]$, $[\ell,\infty)$, $[-\infty,\ell]$, $(-\infty,\ell]$, and $[\ell,\ell]=\{\ell\}$ for $\ell\in\bR$. A pair of risk measures $(\phi,\rho):\X\to I(\bR)\times\bR$ is called a \emph{Bayes pair} \citep{EMWW21} if for some \emph{loss function} $S:\bR^2\to \bR$,
$$\phi(X)=\argmin_{a\in\bR}\E[S(a,X)],~\mbox{and}~\rho(X)=\min_{a\in\bR}\E[S(a,X)],~~X\in\X.$$
If $\phi$ is further cash additive (naturally defined for interval-valued functions), then we call $\rho$ a \emph{Bayes risk measure}, and $\phi$ the corresponding \emph{Bayes estimator}.
It is clear that $(\VaR,\ES)$ is a Bayes pair by \eqref{eq:ES2}. A natural question is whether we can also write $\ES_\Lambda$ as the minimum of some loss function and find its corresponding minimizer.
Ideally, we may expect to find the relation between $\VaR_\Lambda$ and $\ES_\Lambda$ similar to $(\VaR,\ES)$ in optimization.
Below we show a representation of $\ES_\Lambda$ based on the relation \eqref{eq:ES2}, which we call the RU formula of $\ES_\Lambda$. Define the mapping $T_\Lambda:\R\times\R\times L^\infty\to (-\infty,\infty]$ as
\begin{equation}\label{eq:mapping}
    T_\Lambda:(a,x,X)\mapsto \E\left[a+\frac{1}{1-\Lambda(x)}(X-a)_+\right]\vee x.
\end{equation}
The next result also demonstrates the convexity of $T_\Lambda$ in different variables. We use the term ``joint convexity" when there are multiple variables to emphasize its difference from marginal convexity.

\begin{theorem}\label{prop:RU-relation}
    Let $\Lambda:\R\to[0,1]$ be a right-continuous decreasing function  and $T_\Lambda$ be given in \eqref{eq:mapping}. We have
  \begin{equation}\label{eq:RU}
      \ES_\Lambda(X)=\min_{(a,x)\in\R^2}T_\Lambda(a,x,X)=\min_{(a,x)\in\R^2}\left\{\E\left[a+\frac{1}{1-\Lambda(x)}(X-a)_+\right]\vee x\right\},~~X\in L^\infty,
  \end{equation}
    where the minima are obtained at $x^*=\ES_\Lambda(X)$ and $$a^*\left\{\begin{array}{ll}
        \in[\VaR_{\Lambda(x^*)}(X),\VaR^+_{\Lambda(x^*)}(X)], & \mbox{if}~ \Lambda(x^*)\in[0,1),\\
         =\VaR_{1}(X), & \mbox{if}~ \Lambda(x^*)=1. 
    \end{array}\right.$$
    Moreover, 
    \begin{enumerate}[(i)]
        \item $T_\Lambda(a,x,X)$ is jointly convex in $(a,X)\in \R\times L^\infty$ for all $x\in\R$;
        \item  $T_\Lambda(a,x,X)$   is convex in $x\in\R$ for all $(a,X)\in \R\times L^\infty$ if and only if the function $x\mapsto 1/(1-\Lambda(x))$ is convex;
        \item the following statements are equivalent: 
        \begin{enumerate}
            \item $T_\Lambda(a,x,X)$  is jointly convex   in $(a,x)\in\R^2$ for all $X\in L^\infty$; 
            \item $T_\Lambda(a,x,X)$  is jointly quasi-convex   in $(a,x)\in\R^2$ for all $X\in L^\infty$; 
            \item  $T_\Lambda(a,x,X)$  is jointly convex in $(a,x,X)\in\R\times\R\times L^\infty$;
            \item  $T_\Lambda(a,x,X)$  is jointly  quasi-convex in $(a,x,X)\in\R\times\R\times L^\infty$;
            \item $\Lambda$ is constant on $\R$.
        \end{enumerate}
    \end{enumerate}
\end{theorem}
\begin{proof}
For any $X\in L^\infty$, we have by Theorem \ref{thm:rep}, formulation \eqref{eq:ES2}, and Remark \ref{rem:cont} that
\begin{align*}
    \ES_\Lambda(X)=\min_{x\in\R}\left\{\ES_{\Lambda(x)}(X)\vee x\right\}&=\min_{x\in\R}\left\{\min_{a\in\R}\E\left[a+\frac{1}{1-\Lambda(x)}(X-a)_+\right]\vee x\right\}\\
    &=\min_{x\in\R}\min_{a\in\R}\left\{\E\left[a+\frac{1}{1-\Lambda(x)}(X-a)_+\right]\vee x\right\}.
    % &=\min_{a\in\R}\min_{x\in\R}\left\{\E\left[a+\frac{1}{1-\Lambda(x)}(X-a)_+\right]\vee x\right\}.
\end{align*}
For the optimization problem above, the minimizer $x^*=\ES_\Lambda(X)$ is obtained by definition \eqref{eq:gate}, and the minimizer $a^*$ is obtained by \eqref{eq:VaR2}.

Statement (i) is straightforward. We prove statements (ii) - (iv).

    (ii) The ``if" part is clear by the convexity of $x\mapsto 1/(1-\Lambda(x))$. To show the ``only if" part, suppose \eqref{eq:mapping} is convex in $x$. Let
    $\bar x=\inf\{x\in\R:\Lambda(x)=0\}.$
    Right-continuity of $\Lambda$ yields that $\Lambda(\bar x)=0$.
    We first prove $x\mapsto 1/(1-\Lambda(x))$ is convex in $x\in(-\infty,\bar x)$.
    Suppose for contradiction that
    \begin{equation}\label{eq:counter1}
        \frac{1}{1-\Lambda\left(\frac{x_0+y_0}{2}\right)}>\frac{1}{2(1-\Lambda(x_0))}+\frac{1}{2(1-\Lambda(y_0))},~~\mbox{for some}~x_0,y_0\in(-\infty,\bar x).
    \end{equation}
    Because it is clear that 
    \begin{equation}\label{eq:counter3}
        \lim_{a\downarrow -\infty}\E\left[a+\frac{1}{1-\Lambda(x)}(X-a)_+\right]=\infty,~~\mbox{for all}~x\in(-\infty,\bar x),
    \end{equation}
    there exists $a_0\in\R$, such that
    \begin{equation}\label{eq:counter2}
        \begin{aligned}
        &\E\left[a_0+\frac{1}{1-\Lambda(x_0)}(X-a_0)_+\right]\ge x_0,~\E\left[a_0+\frac{1}{1-\Lambda(y_0)}(X-a_0)_+\right]\ge y_0,\\
    \mbox{and}~&\E\left[a_0+\frac{1}{1-\Lambda\left((x_0+y_0)/2\right)}(X-a_0)_+\right]\ge \frac{x_0+y_0}{2}.
    \end{aligned}
    \end{equation}
    \eqref{eq:counter1} and \eqref{eq:counter2} together contradict the fact that \eqref{eq:mapping} is convex in $x$. Therefore, the function $x\mapsto 1/(1-\Lambda(x))$ is convex in $x\in(-\infty,\bar x)$.

    Next, we show that $x\mapsto 1/(1-\Lambda(x))$ is convex in $x\in\R$. Because $x\mapsto 1/(1-\Lambda(x))$ is decreasing, it suffices to show that $x\mapsto 1/(1-\Lambda(x))$ is continuous at $\bar x$ if $\bar x<\infty$. Suppose for contradiction that $\Lambda(\bar x-)>0$. Because of \eqref{eq:counter3}, there exists $a_1\in(-\infty,\esssup(X))$, such that
    $
     % \label{eq:counter4}
        \E[a_1+(X-a_1)_+/(1-\Lambda(\bar x-))]>\bar x.
    $
    It is clear that $\E[(X-a_1)_+]>0$ and thus
    $$\E\left[a_1+\frac{1}{1-\Lambda(\bar x-)}(X-a_1)_+\right]>\E\left[a_1+(X-a_1)_+\right].$$
    It follows that
    \begin{equation*}
    % \label{eq:counter5}
    \begin{aligned}
        &\lim_{x\uparrow\bar x}\frac{\E\left[a_1+\frac{1}{1-\Lambda(x)}(X-a_1)_+\right]\vee x}{2}+\frac{\E\left[a_1+(X-a_1)_+\right]\vee \bar x}{2}\\
        % &=\frac{\E\left[a_1+\frac{1}{1-\Lambda(\bar x-)}(X-a_1)_+\right]}{2}+\frac{\E\left[a_1+(X-a_1)_+\right]\vee \bar x}{2}\\
        &<\E\left[a_1+\frac{1}{1-\Lambda(\bar x-)}(X-a_1)_+\right]=\lim_{x\uparrow \bar x}\left\{\E\left[a_1+\frac{1}{1-\Lambda((x+\bar x)/2)}(X-a_1)_+\right]\vee \frac{x+\bar x}{2}\right\}.
    \end{aligned}
    \end{equation*}
    This contradicts with the convexity of \eqref{eq:mapping}. Therefore, $x\mapsto 1/(1-\Lambda(x))$ is convex in $x\in\R$.

    Next, we prove statement (iii). It is straightforward that (a) $\Rightarrow$ (b) and (c) $\Rightarrow$ (d).
    
    ``(e) $\Rightarrow$ (a)": This follows by the convexity of \eqref{eq:mapping} in $a\in\R$ and the fact that an increasing convex transform of a convex function is still convex. We show the ``only if" part.
    
    ``(b) $\Rightarrow$ (e)": Suppose that \eqref{eq:mapping} is jointly quasi-convex in $(a,x)\in\R^2$ for all $X\in L^\infty$. Suppose for contradiction that $\Lambda$ is decreasing and non-constant on $\R$. There exists $x,y,t\in\R$ with $y<x\le t$, such that $\Lambda(y)\ge\Lambda((x+y)/2)>\Lambda(x)$. Take $a<b=t$, and $X\in L^\infty$ with $\p(X=a)=1-\p(X=t)=\Lambda(x)$. Because $\Lambda(x)<1$, we have
    $$\E\left[a+\frac{1}{1-\Lambda(x)}(X-a)_+\right]\vee x=\E\left[b+\frac{1}{1-\Lambda(y)}(X-b)_+\right]\vee y=t,$$
    whereas
    $$\begin{aligned}
        &\E\left[\frac{a+b}{2}+\frac{1}{1-\Lambda\left(\frac{x+y}{2}\right)}\left(X-\frac{a+b}{2}\right)_+\right]\vee \frac{x+y}{2}\\
        &=\left\{\frac{a+t}{2}+\frac{1-\Lambda(x)}{1-\Lambda\left(\frac{x+y}{2}\right)}\left(t-\frac{a+t}{2}\right)\right\}\vee \frac{x+y}{2}>t.
    \end{aligned}$$
    This contradicts the joint quasi-convexity of \eqref{eq:mapping} in $(a,x)\in\R^2$, and thus $\Lambda$ is constant on $\R$.

    ``(e) $\Rightarrow$ (c)": This follows by statement (i) and the fact that an increasing convex transform of a convex function is still convex.
    
    ``(d) $\Rightarrow$ (e)": This follows directly by the proof for the implication ``(b) $\Rightarrow$ (e)". 
\end{proof}

% \textcolor{red}{
% [QW: I don't think the function $(a,x,X)\mapsto\E\left[a+\frac{1}{1-\Lambda(x)}(X-a)_+\right]\vee x$ can be jointly quasi-convex whenever $\Lambda$ is non-constant on $(0,1)$. A counterexample (even when $X$ is deterministic) can be the following: 
% The best I can do is to prove that $(a,x,X)\mapsto\E\left[a+\frac{1}{1-\Lambda(x)}(X-a)_+\right]\vee x$ is jointly quasi-convex in $(x,X)$ if $\Lambda$ is convex (maybe if and only if). Moreover, the function $(a,x,X)\mapsto\E\left[\frac{1}{1-\Lambda(x+a)}(X-a)_+\right]\vee x$ is jointly quasi-convex in $(a,x,X)$ if $\Lambda$ is convex, where the term ``$+a$" is dropped.]
% }

For the RU formula of $\Lambda$-ES, Theorem \ref{prop:RU-relation} indicates that we do not guarantee joint convexity (or quasi-convexity) of the objective \eqref{eq:mapping} in $(a,x)\in\R^2$ or $(a,x,X)\in\R\times\R\times L^\infty$ unless $\Lambda$ is a constant (i.e., $\ES_\Lambda$ is an ES). Nevertheless, the mapping \eqref{eq:mapping} is convex in $x\in\R$ when the function $x\mapsto 1/(1-\Lambda(x))$ is convex.
Two examples satisfying this condition are: $\Lambda$ is a constant (i.e.~the usual ES)
and  $\Lambda (x) = (e^{ax}+1)^{-1}$ for $a>0$.\footnote{The example $\Lambda(x)=(e^{ax}+1)^{-1}$, $a>0$, $x\in\R$, provides a suggestion of a non-trivial choice of the $\ES_\Lambda$ risk measure to use in practice. By Theorem \ref{prop:RU-relation}, an obvious advantage of such a choice is that it makes the objective of a practical $\Lambda$-ES-based optimization problem convex in the variable $x$.}
Moreover, Theorem \ref{prop:RU-relation} shows that $\ES_\Lambda$ has a similar feature to a  Bayes risk measure introduced by \cite{EMWW21}, as it can be represented as the minimum of an expected loss function with an additional transformation.
% \footnote{Let $I(\bR)$ be the set of intervals on $\bR$. A risk measure $\rho:\X\to\bR$ is called a \emph{Bayes risk measure} if $$\phi(X)=\argmin_{a\in\bR}\E[S(a,X)]~\mbox{and}~\rho(X)=\min_{a\in\bR}\E[S(a,X)],~~X\in\X,$$
% where the risk measure $\phi:\X\to I(\bR)$ is translation invariant.}
% The term inside the minimization in \eqref{eq:mapping} is jointly convex in $(X,a)$. It is also quasi-convex in $x$, but not necessarily convex in $x$, depending on the choice of $\Lambda$. \com{I think a statement of convexity can be included in Proposition 3, and it can be even made into a theorem if there is something nontrivial to verify. For instance, can we prove that it is  jointly quasi-convex in $(X,a,x)$? Under what conditions it is jointly convex in $(X,a,x)$?} 
The corresponding minimizer is the interval of the left- and right-quantiles at the level of $\Lambda(x^*)$  instead of $[\VaR_\Lambda,\VaR^+_\Lambda]$.
%It remains an open question whether $\ES_\Lambda$ is a Bayes risk measure.

A possible direction to explore the issue of Bayes pair is   through the scoring function of $\Lambda$-$\VaR$.\footnote{A set-valued functional $\rho:\X\to 2^\R$ is called \emph{elicitable} on $\X$ if there exists a function (\emph{scoring function}) $S:\bR^2\to\bR$, such that $$\rho(X)=\argmin_{a\in\bR}\E[S(a,X)],~~X\in\X.$$} \cite{BB15} and \cite{BPR17} showed that for a decreasing function $\Lambda:\R\to(0,1)$, $\VaR_\Lambda$ is elicitable with the scoring function
$$
S_\Lambda(a,y)= (a-y)_+ - \int_y^a \Lambda (t) \,\d t = (y-a)_+ - \int_{a}^y (1-\Lambda(t))\,\d t,~~a\in\bR,~y\in\R.$$
The pair of risk measures we get with the above scoring function is $(\VaR_\Lambda,\rho)$, where
 $$
\rho_\Lambda(X) = \min_{a\in \bR}\E[c S_\Lambda(a,X)+f(X)],~~X\in L^\infty,
 $$
for some constant $c>0$ and real function $f:\R\to\R$. The risk measure $(\VaR_\Lambda,\rho_\Lambda)$ is not a Bayes pair because $\VaR_\Lambda$ is not cash additive in general. Moreover, we find that $\rho_\Lambda$ cannot satisfy quasi-convexity, normalization, and $\rho_\Lambda\ge \VaR_\Lambda$ simultaneously, and thus does not coincide with $\ES_\Lambda$ for any choices of $c$ and $f$. We put the detailed arguments for this conflict in Appendix \ref{sec:others}.

\subsection{Optimization with Lambda ES}

 We briefly discuss optimization problems with $\Lambda$-ES as a constraint or an objective, which turn out to be closely related to corresponding problems for ES.  %Let $\Lambda:\R\to[0,1]$ be a decreasing function. 
 For $n\in \N$, let $\mathbf{L}=(L_1,\dots,L_n)\in (L^\infty)^n$ represent a vector of losses, $\bm{\theta} \in \Theta$ represent a decision variable, where $\Theta$ is a convex set of actions, and $f: \Theta\times \R^n \to \R$ represent a loss function. In a typical example in finance,  $\mathbf{L}$ represents losses from multiple assets and $\bm{\theta}$ represents a portfolio choice.

First, we are interested in the setting   where the decision maker minimizes an objective risk measure $\rho:L^\infty\to\bR$ of the aggregate loss $f(\bm{\theta},\mathbf{L})$, guaranteeing that the $\Lambda$-ES of the total loss does not exceed a pre-specified value $\ell \in\R$. Namely, we consider the following optimization problem:
\begin{equation}\label{eq:prob_con}
\min_{\bm{\theta}\in\Theta}\rho(f(\bm{\theta},\mathbf{L}))~~~\mbox{subject to}~\ES_\Lambda(f(\bm{\theta},\mathbf{L}))\le \ell.
\end{equation}
It is straightforward to show that the constraint in \eqref{eq:prob_con} is equivalent an ES constraint under a mild condition.
\begin{proposition}\label{prop:prob_con}
    Let $\Lambda:\R\to[0,1]$ be a right-continuous decreasing function. The constraint $\ES_\Lambda(f(\bm{\theta},\mathbf{L}))\le \ell$ in \eqref{eq:prob_con} is equivalent to $\ES_{\Lambda(\ell)}(f(\bm{\theta},\mathbf{L}))\le \ell$.
\end{proposition}
\begin{proof}
    By definition, we have
    \begin{align*}
        \ES_\Lambda(f(\bm{\theta},\mathbf{L}))\le \ell &\iff\sup_{x\in\R}\left(\ES_{\Lambda(x)}(f(\bm{\theta},\mathbf{L}))\wedge x\right)\le \ell\\
        % &\iff \mbox{for all }x\in\R:~\ES_{\Lambda(x)}(f(\bm{\theta},\mathbf{L}))\wedge x\le \ell\\
        &\iff \mbox{for all }x\in\R:~\ES_{\Lambda(x)}(f(\bm{\theta},\mathbf{L}))\le \ell~\mbox{or}~x\le \ell\\
        &\iff \sup_{x>\ell }\ES_{\Lambda(x)}(f(\bm{\theta},\mathbf{L}))\le \ell\iff \ES_{\Lambda(\ell)}(f(\bm{\theta},\mathbf{L}))\le \ell,
    \end{align*}
    where the last equivalence holds by right-continuity of $\Lambda$.
\end{proof}
%Proposition \ref{prop:prob_con} implies that a $\Lambda$-ES-constrained optimization problem of level $\ell\in\R$ can be equivalently converted to a problem with $\ES_{\Lambda(\ell)}$ constrained below the same level $\ell$.
ES-constrained optimization problems have been studied extensively in finance and operations research (see e.g., \citealp{RU02, KPU02,ZKR13}). 

A related optimization problem is to minimize $\Lambda$-ES as an objective:
\begin{equation}\label{eq:prob_obj}
    \min_{\bm{\theta}\in\Theta}\ES_{\Lambda}(f(\bm{\theta},\mathbf{L})).
\end{equation}
Problems in the form of \eqref{eq:prob_obj} for ES are also well studied in operations research \citep[see e.g.,][]{ZF09,CLM25,WLM25}.
The result below provides general insights into solving \eqref{eq:prob_obj}.   Define the mapping $\widetilde T^f_\Lambda:\Theta\times\R\times\R\times (L^\infty)^n\to\bR$ as
\begin{equation}\label{eq:mapping_portfolio}
    \widetilde T^f_\Lambda(\bm\theta,a,x,\mathbf{L})=\E\left[a+\frac{1}{1-\Lambda(x)}(f(\bm{\theta},\mathbf{L})-a)_+\right]\vee x.
\end{equation}
% We have the following result.

\begin{proposition}\label{prop:prob_obj}
    For a right-continuous decreasing function $\Lambda:\R\to[0,1]$ and $\widetilde T^f_\Lambda  $ defined in \eqref{eq:mapping_portfolio},
    \begin{equation}\label{eq:prob_portfolio}
        \min_{\bm{\theta}\in\Theta}\ES_{\Lambda}(f(\bm{\theta},\mathbf{L}))=\min_{(\bm{\theta},a,x)\in\Theta \times\R\times \R} \widetilde T^f_\Lambda(\bm\theta,a,x,\mathbf{L}) .
    \end{equation}
    Moreover,  
    \begin{enumerate}[(i)]
        \item $\widetilde T^f_\Lambda(\bm\theta,a,x,\mathbf{L})$ is convex in $a\in\R$ for all $(\bm\theta,x,\mathbf{L})\in\Theta\times\R\times (L^\infty)^n$.
        \item if in addition, $\Lambda$ is not constantly $1$, then $\widetilde T^f_\Lambda(\bm\theta,a,x,\mathbf{L})$ is jointly convex in $(\bm\theta, \mathbf{L})\in\Theta \times (L^\infty)^n$ for all $(a,x)\in\R^2$ if and only if $f(\bm\theta,\mathbf{L})$ is jointly convex in $(\bm\theta,\mathbf{L})$.
        \item $\widetilde T^f_\Lambda(\bm\theta,a,x,\mathbf{L})$ is convex in $x\in\R$ for all $(\bm\theta,a,\mathbf{L})\in\Theta\times\R\times (L^\infty)^n$ if and only if the function $x\mapsto 1/(1-\Lambda(x))$ is convex.
    \end{enumerate}
\end{proposition}
\begin{proof}
    Equation \eqref{eq:prob_portfolio} and statements (i) and (iii) follow from Theorem \ref{prop:RU-relation}. Statement (ii) is straightforward to verify. 
\end{proof}

Although the results in Propositions \ref{prop:prob_con}--\ref{prop:prob_obj} are quite simple, they illustrate that $\ES_\Lambda$-based optimization problems share many similar features to ES-based ones, which are widely studied.

% As briefly mentioned  in Section \ref{sec:5RU}, the optimization problem in Proposition \ref{prop:prob_obj} has some forms of convexity. When $f$ is convex, the minimization problem in Proposition \ref{prop:prob_obj} is convex in $\bm\theta$, $\mathbf L$ and $a$, but not necessarily in $x$. To obtain a convex optimization problem, a sufficient condition is that $x\mapsto 1/(1-\Lambda(x))$ is convex.
% \com{Question: related to the comment after Prop 3, is this a if and only if statement? If so, we can write it as a simple result. We can even promote $\Lambda (x) = (e^{ax}+1)^{-1}$ as a new $\ES_\Lambda$ risk measure to use in practice.
% Not sure whether convexity of $x\mapsto 1/(1-\Lambda(x))$ implies joint convexity in $(\bm \theta,\mathbf L, a,x)$. If so it would be nice.}  

% \textcolor{red}{QW: We can possibly combine this section with Section 6 if we still need more applications.}

% \section{Application}

\section{Extensions to the space of integrable random variabls}

\label{sec:L1}

In this section, we extend our discussions on ES and $\Lambda$-ES from the space of $L^\infty$ to $L^1$.
Similarly to the corresponding definitions on $L^\infty$, we define ES at level $\alpha \in [0,1]$ as the mapping 
$\ES_\alpha:L^1\to \bR$  given by
$$
% \begin{align}
% \label{eq:ES1-L1}
\ES_\alpha(X)=\frac{1}{1-\alpha} \int_\alpha^1 \VaR_\beta (X) \,\d \beta  \mbox{ for $\alpha\in [0,1)$}, 
% \end{align}
$$
and $\ES_1(X)=\VaR_1(X)$; for a decreasing function $\Lambda:\R\to[0,1]$, we define $\ES_\Lambda:L^1\to\bR$ as
\begin{equation*}
    \ES_\Lambda(X)=\sup_{x\in\R}\left\{\ES_{\Lambda(x)}(X)\wedge x\right\},~~X\in L^1.
\end{equation*}

% $$
% \min\left\{\rho:L^1\to\bR~|~\rho\ge\VaR_\Lambda~\mbox{and}~\rho~\mbox{is quasi-convex and law-invariant}\right\}.
% $$

Some of the results in the previous sections can be naturally extended to $L^1$, whereas others only hold under a weakened setup, for which we provide independent proofs for completeness. For the convenience of our discussion, we first note that the properties in Propositions \ref{prop:properties} and \ref{prop:convex} still hold for $\ES_\Lambda$ on $L^1$ by the same arguments in its proof replacing $L^\infty$ by $L^1$.

\subsection{Finiteness of Lambda ES}

Below we show the finiteness of $\Lambda$-ES on $L^1$. As a result, the risk measure $\ES_\Lambda$ is always well-defined (possibly being $\infty$) on $L^1$.
\begin{proposition}
Let $\Lambda:\R\to[0,1]$ be a decreasing function.
The mapping $\ES_\Lambda:L^1\to \bR$ 
satisfies
$$-\infty<\E[X] \le \ES_\Lambda(X) \le \ES_1(X),~~~X\in L^1.
$$
In particular, $\ES_\Lambda(X)$ is finite on $L^1$ if and only if $\VaR_1(X)<\infty$ or
$\Lambda
$ is not constantly $1$.
\end{proposition}
\begin{proof}
The relation $-\infty<\E \le \ES_\Lambda \le \ES_1$ holds by item (a) of Proposition \ref{prop:properties} on $L^1$. We prove the ``if" part of the second statement, whose ``only if" part is straightforward.

For any $X\in L^1$, first, suppose that $\VaR_1(X)<\infty$. It is straightforward that $\ES_\Lambda(X)\le\ES_1(X)=\VaR_1(X)<\infty$. Next, suppose that $\Lambda$ is not constantly $1$. There exists $x_0\in\R$, such that $0\le\Lambda(x_0)<1$. It follows that $\ES_{\Lambda(x_0)}(X)<\infty$. By \eqref{eq:gate1},
$$\ES_\Lambda(X)=\inf_{x\in\R}\left(\ES_{\Lambda(x)}(X)\vee x\right)\le \ES_{\Lambda(x_0)}(X)\vee x_0<\infty.$$
The proof is complete.
\end{proof}

\subsection{Dominance of ES and Lambda ES}

Here, we examine the $L^1$ versions of the dominance results in Theorems \ref{lem:ES-domin} and \ref{thm:rep}.
Theorem \ref{lem:ES-domin} does not hold in general if we extend the space of $\ES_\alpha$ from $L^\infty$ to $L^1$ because the dominance may fail at $\alpha=0$. A counterexample can be: Let $\rho:L^1\to\bR$ be a risk measure defined as follows.
$$\rho(X)=\left\{\begin{array}{ll}
    \esssup(X), & \mbox{if }\esssup(X)=\infty~\mbox{or}~\essinf(X)>-\infty,\\
    -\infty, & \mbox{if }\esssup(X)<\infty~\mbox{and}~\essinf(X)=-\infty,
\end{array}\right.~X\in L^1.$$
One can check that $\rho$ is quasi-convex, law invariant, and $\rho\ge \VaR^+_0$. However, the condition $\rho\ge \E$ fails. Therefore, $\E$ is not the smallest quasi-convex and law-invariant risk measure dominating $\VaR^+_0$ and thus \eqref{eq:ES3-prime} in Theorem \ref{lem:ES-domin} fails for $\alpha=0$.

Below, we state the dominance results for ES over VaR and $\Lambda$-ES over $\Lambda$-VaR on the space of $L^1$. Both results rely on slightly stronger assumptions than Theorems \ref{lem:ES-domin} and \ref{thm:rep} regarding the case of $\alpha=0$. Write $\underline{L}^1$ as the set of all random variables in $L^1$ that are essentially bounded  from below.

\begin{theorem}
    \label{thm:ES-domin-L1}
For any $\alpha \in (0,1]$,
\begin{align}
\label{eq:ES3-L1}
\ES_\alpha = \min \{\rho:L^1 \to \bR \mid \rho\ge \VaR_\alpha~\mbox{and $\rho$ is quasi-convex and law invariant}\}. 
\end{align} 
For any $\alpha \in (0,1)$,
\begin{align}
\label{eq:ES3-prime-L1}
\ES_\alpha = \min \{\rho:L^1 \to \bR \mid \rho\ge \VaR^+_\alpha~\mbox{and $\rho$ is quasi-convex and law invariant}\}.
\end{align}
For any $\alpha \in [0,1)$,
\begin{align}
\label{eq:ES3-prime-L1-bb}
\ES_\alpha = \min \{\rho:\underline{L}^1 \to \bR \mid \rho\ge \VaR^+_\alpha~\mbox{and $\rho$ is quasi-convex and law invariant}\}.
\end{align} 
\end{theorem}
\begin{proof}
    The proofs of \eqref{eq:ES3-L1},  \eqref{eq:ES3-prime-L1}, and the case of $\alpha\in(0,1)$ in \eqref{eq:ES3-prime-L1-bb} follow directly from those for Theorem \ref{lem:ES-domin} whose arguments still hold on $L^1$. We only need to prove \eqref{eq:ES3-prime-L1-bb} for $\alpha=0$.

    For any $n\in\N$ and $X\in \underline{L}^1$ with distribution $F$, write $K_n=\sqrt{n}+\VaR^+_0(X)$. By Corollary A.3 of \cite{EWW15}, there exist $\widetilde X_i\lawis F$, $i\in[n]$, such that
    $$\left|\frac{1}{n}\sum^n_{i=1}\left(\widetilde X_i\wedge K_n\right)-\E\left[X\wedge K_n\right]\right|\le \frac{\sqrt{n}}{n}=\frac{1}{\sqrt{n}}.$$
    It follows that
    $$\frac{1}{n}\sum^n_{i=1}\left(\widetilde X_i\wedge K_n\right)\ge \E\left[X\wedge K_n\right]-\frac{1}{\sqrt{n}},$$
    which implies that
    $$\E\left[X\wedge K_n\right]\ge\frac{1}{n}\VaR^+_0\left(\left(\widetilde X_1\wedge K_n\right)+\cdots+\left(\widetilde X_n\wedge K_n\right)\right)\ge \E\left[X\wedge K_n\right]-\frac{1}{\sqrt{n}}.$$
    Letting $n\to\infty$, by monotone convergence theorem, we have
    $$\frac{1}{n}\sup\left\{\VaR^+_0\left({X}_1+\cdots+{X}_n\right):X_i\lawis F,~i\in[n]\right\}\to\E[X].$$
    The rest of the proof follows from that for Theorem \ref{lem:ES-domin}.
\end{proof}

With Theorem \ref{thm:ES-domin-L1}, we obtain 
 the following result, which extends the domain in Theorem \ref{thm:rep} from $L^\infty$  to $L^1$ or $\underline{L}^1$.

\begin{theorem}\label{thm:rep-L1}
The following statements hold.
\begin{enumerate}[(i)]
    \item For a decreasing function $\Lambda:\R\to(0,1]$, the smallest quasi-convex and law-invariant risk measure on $L^1$ dominating $\VaR_\Lambda$ is $\ES_\Lambda$, that is, 
  $$
    \ES_\Lambda = \min \{\rho : L^1\to \bR \mid \rho\ge \VaR_\Lambda~\mbox{and $\rho$ is quasi-convex and law invariant}\}.
     $$
     \item For a decreasing function $\Lambda:\R\to(0,1)$, the smallest quasi-convex and law-invariant risk measure on $L^1$ dominating $\VaR^+_\Lambda$ is $\ES_\Lambda$, that is, 
  $$
    \ES_\Lambda = \min \{\rho : L^1\to \bR \mid \rho\ge \VaR^+_\Lambda~\mbox{and $\rho$ is quasi-convex and law invariant}\}.
     $$
     \item For a decreasing function $\Lambda:\R\to[0,1)$, the smallest quasi-convex and law-invariant risk measure on $\underline{L}^1$ dominating $\VaR^+_\Lambda$ is $\ES_\Lambda$, that is, 
  $$
    \ES_\Lambda = \min \{\rho : \underline{L}^1\to \bR \mid \rho\ge \VaR^+_\Lambda~\mbox{and $\rho$ is quasi-convex and law invariant}\}.
     $$
\end{enumerate}
Moreover, the identity holds for all decreasing functions $\Lambda:\R\to[0,1]$:
$$
% \begin{align}
% \label{eq:gate1-L1}
 %\sup_{x\in \R} \left(\ES_{\Lambda(x)}(X) \wedge x\right)= 
 \ES_{\Lambda}(X)  = \inf_{x\in \R} \left( \ES_{\Lambda(x)}(X) \vee x \right),~~~X\in L^1.
% \end{align} 
$$
\end{theorem}

\subsection{Dual representation and RU formula for Lambda ES}

In this section, we restate the dual representation (Theorem \ref{thm:dual}) and the RU formula (Theorem \ref{prop:RU-relation}) for $\Lambda$-ES on $L^1$. The proofs of both results below follow from the same arguments as those for Theorems \ref{thm:dual} and \ref{prop:RU-relation} by replacing $L^\infty$ with $L^1$.
\begin{theorem}\label{thm:dual-L1}
     For any decreasing function $\Lambda:\R\to[0,1]$, the risk measure $\ES_\Lambda$ adopts the following representation:
     \begin{equation}\label{eq:dual-L1}
         \ES_\Lambda(X)=\sup_{\q\in\M_{1,f}} R(\E_\q [X],\q), ~~~X\in L^1,
     \end{equation}
     where for $(t,\q)\in \R\times \mathcal M_{1,f}$,
 $$
        R(t,\q)=\sup_{x\in\R}\left\{t\wedge x~:~\Lambda(x)\ge 1-\frac{\mathrm{d}\p}{\mathrm{d}\q},~\q\mbox{-almost surely}\right\}.
 $$
     Moreover, the following are true:
     \begin{enumerate}[(i)]
        \item The supremum in \eqref{eq:dual-L1} can be changed to a maximum if $\Lambda$ is left-continuous.
         \item $(t,\q)\mapsto R(t,\q)$ is upper semicontinuous, quasi-concave, and increasing in $t$;
         \item $\inf_{t\in\R} R(t,\q)=\inf_{t\in\R} R(t,\q^\prime)$ for all $\q,\q^\prime\in \M_{1,f}$;
         \item $R(t_1,\q)-R(t_2,\q)\le t_1-t_2$ for all $t_1\ge t_2$ and $\q\in\M_{1,f}$.
     \end{enumerate}
\end{theorem}

With a slight abuse of notation, we define the mapping $T_\Lambda:\bR\times\R\times L^1\to\bR$ as
\begin{equation}\label{eq:mapping-L1}
    T_\Lambda:(a,x,X)\mapsto \E\left[a+\frac{1}{1-\Lambda(x)}(X-a)_+\right]\vee x.
\end{equation}

\begin{proposition}\label{prop:RU-relation-L1}
    For any right-continuous decreasing function $\Lambda:\R\to[0,1]$, the risk measure $\ES_\Lambda$ can be represented as follows.
 $$   \ES_\Lambda(X)=\min_{(a,x)\in\bR\times\R}T(a,x,X)=\min_{(a,x)\in\bR\times\R}\left\{\E\left[a+\frac{1}{1-\Lambda(x)}(X-a)_+\right]\vee x\right\},~~X\in L^1,
 $$
    where the minima are obtained at $x^*=\ES_\Lambda(X)$ and $$a^*\left\{\begin{array}{ll}
        \in[\VaR_{\Lambda(x^*)}(X),\VaR^+_{\Lambda(x^*)}(X)], & \mbox{if}~ \Lambda(x^*)\in[0,1),\\
         =\VaR_{1}(X), & \mbox{if}~ \Lambda(x^*)=1. 
    \end{array}\right.$$
    Moreover, for $T_\Lambda$ defined in \eqref{eq:mapping-L1}, 
    \begin{enumerate}[(i)]
        \item $T_\Lambda(a,x,X)$ is jointly convex in $(a,X)\in \bR\times L^1$ for all $x\in\R$;
        \item  $T_\Lambda(a,x,X)$   is convex in $x\in\R$ for all $(a,X)\in \bR\times L^1$ if and only if the function $x\mapsto 1/(1-\Lambda(x))$ is convex;
        \item the following statements are equivalent: 
        \begin{enumerate}
            \item $T_\Lambda(a,x,X)$  is jointly convex   in $(a,x)\in\bR\times\R$ for all $X\in L^1$; 
            \item $T_\Lambda(a,x,X)$  is jointly quasi-convex   in $(a,x)\in\bR\times\R$ for all $X\in L^1$; 
            \item  $T_\Lambda(a,x,X)$  is jointly convex in $(a,x,X)\in\bR\times\R\times L^1$;
            \item  $T_\Lambda(a,x,X)$  is jointly  quasi-convex in $(a,x,X)\in\bR\times\R\times L^1$;
            \item $\Lambda$ is constant on $\R$.
        \end{enumerate}
    \end{enumerate}
\end{proposition}

\section{Conclusion}
\label{sec:con}

This paper introduces the $\Lambda$-ES, a new and theoretically grounded generalization of ES that robustly extends the $\Lambda$-VaR framework.
% Our core contribution is the explicit and rigorously justified representation of $\Lambda$-ES, positioning it as the natural and canonical ``ES counterpart" to $\Lambda$-VaR.
We obtain an explicit representation of $\Lambda$-ES and verify that it satisfies several crucial properties as a desired counterpart to $\Lambda$-VaR. In particular, we show that $\Lambda$-ES is the smallest quasi-convex and law-invariant risk measure that dominates $\Lambda$-VaR. The dual representation of $\Lambda$-ES further connects it to established results on quasi-convex cash-subadditive risk measures. The RU formula of $\Lambda$-ES provides useful insights for its potential applications to optimization problems.
% From a theoretical perspective, our most significant insight is that $\Lambda$-ES is not merely a new definition, but the smallest quasi-convex and law-invariant risk measure that dominates $\Lambda$-VaR. This property is paramount, as it establishes $\Lambda$-ES as the most parsimonious construction that satisfies fundamental axioms of prudence and consistency while coherently bounding $\Lambda$-VaR. This mathematical optimality mirrors the foundational relationship between ES and VaR, providing a deep structural guarantee for $\Lambda$-ES's desirable characteristics and cementing its role as a benchmark for generalized tail risk measures.
% The rigorous dual representation further enriches this theoretical understanding, connecting $\Lambda$-ES to the broader landscape of quasi-convex risk measures, offering pathways for advanced analytical techniques and economic interpretations.
Practically, $\Lambda$-ES shares the advantages of $\Lambda$-VaR as a flexible model for risk assessment, and it has additional benefits in risk management such as quasi-convexity,  $L^1$-continuity, and dual representation, thus sharing the advantages of ES over VaR.

 \subsection*{Acknowledgments}
% We thank ...... for helpful comments on the paper. 
RW acknowledges financial support from the Natural Sciences and Engineering Research Council (NSERC) of Canada (Grant Nos.~RGPIN-2024-03728, CRC-2022-00141).

% \newpage

\setcounter{equation}{0}
\renewcommand{\theequation}{\thesection.\arabic{equation}}

\appendix

\section{Other possible formulations of Lambda ES}
\label{sec:others}

% \com{I think this section may be put into the appendix.}
There may be many ways of generalizing ES using a parameter $\Lambda$. 
Below we explain a few possible ways of defining $\Lambda$-ES that fail to satisfy basic requirements, and thus they are not suitable definitions.

\subsection{An algebraic formulation}
We first consider an algebraic way of defining ES.  
We can rewrite  $\ES_\alpha$ as \citep{AT02}
\begin{align*} 
\ES_\alpha(X)=\frac{1}{1-\alpha} \E\left[X\id_{\{ X\ge \VaR_\alpha(X)\}}\right] + \VaR_\alpha(X)(1-\alpha-\p(X\ge \VaR_\alpha(X))).
\end{align*} 
% In particular, if $\p(X\ge \VaR_\alpha(X))=1-\alpha$, then the above formula becomes 
% \begin{align*} 
% \ES_\alpha(X)= \E\left[X \mid  X\ge \VaR_\alpha(X) \right].
% \end{align*}
% This holds in particular when $X$ has a continuous distribution. 
% Inspired by this, we
Denote by $L^1_c$ the set of all random variables in $L^1$ with continuous distributions. We have
\begin{align}
\label{eq:ES4}
\ES_\alpha(X)= \E\left[X \mid  X\ge \VaR_\alpha(X) \right],~~~X\in L^1_c.\end{align}
To define $\Lambda$-ES on continuous random variables using the idea of formulation  \eqref{eq:ES4}, a choice is:
\begin{align}
\label{eq:LES4}
\rho_\Lambda(X)= \E\left[X \mid  X\ge \VaR_\Lambda(X) \right],~~~X\in L^1_c,
\end{align}
for a decreasing function $\Lambda:\R\to (0,1)$.
However, this formulation is not monotone, as the following example shows.
    Let $\Omega = [0,1]$ with $\p$ the Lebesgue measure. Let $0<\epsilon<10$.
    % \com{I rephrased what I meant. Is it necessary to work out how small $\epsilon$ have to be? --- Does it need to be small? If so, make it clear}
    For $\omega \in \Omega$ define
    $$X(\omega) =  \epsilon(\omega-0.1) + \id_{(0.1, 0.9]}(\omega) + 10\cdot  \id_{(0.9, 1]}(\omega)\mbox{~and~}Y(\omega) = \epsilon(\omega-1) + 10\cdot  \id_{(0.9, 1]}(\omega),$$ so that $X \geq Y$. Let $\Lambda$ be a  strictly decreasing function with $\Lambda (1) = 0.1$ and $\Lambda(0) = 0.9$. We can compute  
    $$\VaR_\Lambda(X) = 1\mathrm{~so~that~}\rho_\Lambda(X)= 2+0.45\epsilon.$$ 
    On the other hand, 
    $$\VaR_\Lambda(Y) =  0 \mathrm{~so~that~}\rho_\Lambda(Y)=10-0.05\epsilon.$$
 % Below, we plot the quantile of $X$ in blue, the quantile of $Y$ in green and $\Lambda^{-1}$ in red.
 Taking $\epsilon\downarrow 0$ yields that $\rho_\Lambda$ in \eqref{eq:LES4} is not monotone and is therefore undesirable.

\subsection{A formulation based on the Rockafellar--Uryasev formula}

Another possible formulation of $\Lambda$-ES is based on the RU formula in \eqref{eq:ES2}. Namely, for a decreasing function $\Lambda:\R\to(0,1)$, we may define the following candidate risk measure
\begin{align}
\label{eq:LES2}
\rho_\Lambda(X)= \inf_{x\in \R}\left\{x+\frac{1}{1-\Lambda(x)}\E[(X-x)_+]\right\},~~~X\in L^1.
\end{align}
Here, we use infimum because the minimum may not exist in general.
 Clearly, $\rho_\Lambda$ is monotone in both $\Lambda$ and $X$, is law invariant, and specializes to $\ES_\alpha$ when $\Lambda \equiv \alpha$ for $\alpha \in (0,1)$. Comparing \eqref{eq:LES2} to  Theorem \ref{prop:RU-relation}, we are optimizing \eqref{eq:RU} over the subset $\{(a,x) \in \R^2: a=x\}$ so that the optimum is larger. We have $\ES_\Lambda \leq \rho_\Lambda$ so that $\rho_\Lambda$ also dominates $\VaR_\Lambda$. However,
 the following counterexample shows that $\rho_\Lambda$ defined in \eqref{eq:LES2} is not quasi-convex in general, and is thus not an ideal candidate for $\Lambda$-ES.
 
 % The risk measures $\rho_\Lambda$ and $\ES_\Lambda$ are different. For instance, take the random variable $Y$ from Example \ref{ex:1}, $\rho_\Lambda(Y)$ is close to $2$ but $\ES_\Lambda(Y)<1$.  It is not clear whether $\rho_\Lambda$ is quasi-convex. 
% \com{QW: I think we'd better find a counterexample. Right?}

Let $a_0,b_0,\alpha,\beta,\epsilon\in\R$ with $b_0<a_0$, $\epsilon>0$, and $3/4<\beta<\alpha<1$. Let
$$\Lambda_0(a)=\alpha+(\beta-\alpha)\id_{\{a>a_0\}},~~a\in\R.$$
It is clear that
$$\rho_{\Lambda_0}(X)=\inf_{x\le a_0}\left\{x+\frac{1}{1-\alpha}\E[(X-x)_+]\right\}\wedge \inf_{x>a_0}\left\{x+\frac{1}{1-\beta}\E[(X-x)_+]\right\},~~~X\in L^1.$$
Take $Y,Z\in L^\infty$ such that
$$\p(Z=a_0-\epsilon)=1-\p(Z=b_0)=3/4,~Y=a_0+3\epsilon.$$
It follows that
$$\begin{aligned}
    &\rho_{\Lambda_0}(Z)=\ES_\alpha(Z)=a_0-\epsilon,~\rho_{\Lambda_0}(Y)=a_0+3\epsilon,\\
    \mbox{and}~&\rho_{\Lambda_0}\left(\frac{Y+Z}{2}\right)=a_0+\frac{1}{1-\beta}\E\left[\left(\frac{Y+Z}{2}-a_0\right)_+\right]=a_0+\frac{3/4}{1-\beta}\epsilon>a_0+3\epsilon.
\end{aligned}$$
This indicates that $\rho_{\Lambda_0}$ is not quasi-convex.

\subsection{A formulation based on the score function of Lambda VaR}

Let $\Lambda:\R\to(0,1)$ be a decreasing function.
As discussed in Section \ref{sec:bayes}, a natural possible formulation of $\Lambda$-ES is
$$\rho_\Lambda(X)=\min_{a\in\bR}\E[cS_\Lambda(a,X)+f(X)],~~X\in L^\infty,$$
where $c>0$ is a constant, $f:\R\to\R$ is a real function, and
$$S_\Lambda(a,y)= (a-y)_+ - \int_y^a \Lambda (t) \,\d t = (y-a)_+ - \int_{a}^y (1-\Lambda(t))\,\d t,~~a\in\bR,~y\in\R,$$
is the scoring function for $\VaR_\Lambda$ with
$$\VaR_\Lambda(X)\in\argmin_{a\in\bR}\E[S_\Lambda(a,X)],~~X\in L^\infty.$$
The following argument shows that $\rho_\Lambda$ cannot satisfy quasi-convexity, normalization, and $\rho_\Lambda\ge\VaR_\Lambda$ simultaneously and is thus not a good candidate for $\Lambda$-ES.

\begin{enumerate}[(i)]
    \item Suppose that $\rho_\Lambda$ is normalized. For all $a\in\R$, $\VaR_\Lambda(a)=a$, and thus
$$a=\rho_\Lambda(a)=cS_\Lambda(a,a)+f(a)=f(a).$$
Therefore, we have $f(a)=a$ for all $a\in\R$.

\item Suppose that $\rho_\Lambda$ is normalized and $\rho_\Lambda\ge\VaR_\Lambda$. It implies that $\rho_{\alpha^*}\ge\VaR_{\alpha^*}$ for all $\alpha^*\in [\inf_{x\in\R}\Lambda(x), \sup_{x\in\R}\Lambda(x)]$. For all $X\in L^\infty$,
\begin{align*}
&\E\left[c(X-\VaR_{\alpha^*}(X))_+-c(1-\alpha^*)(X-\VaR_{\alpha^*}(X))+X\right]\\
&=\rho_{\alpha^*}(X)\ge \ES_{\alpha^*}(X) \tag*{\scriptsize \mbox{[Theorem \ref{thm:ES-domin-L1}]}}\\
&=\E\left[\frac{1}{1-\alpha^*}(X-\VaR_{\alpha^*}(X))_+-(X-\VaR_{\alpha^*}(X))+X\right] \tag*{\scriptsize \mbox{[by \eqref{eq:ES2}]}}
\end{align*}
Therefore, $c\ge 1/(1-\alpha^*)$ for all $\alpha^*\in [\inf_{x\in\R}\Lambda(x), \sup_{x\in\R}\Lambda(x)]$, and thus $c\ge 1/(1-\sup_{x\in\R}\Lambda(x))$.

\item Suppose that $\rho_\Lambda$ is quasi-convex, normalized, and $\rho_\Lambda\ge \VaR_\Lambda$.
Let $x_0,y_0,t_0\in\R$, and $\alpha_1,\alpha_2,\alpha_3\in(0,1)$ with
$0<x_0<t_0<y_0$, $\alpha_1<1/4<1/2<\alpha_2<\alpha_3$,
and $$\Lambda_0(x)=\alpha_3\id_{\{x<0\}}+\alpha_2\id_{\{0\le x<t_0\}}+\alpha_1\id_{\{x\ge t_0\}}.$$
Take $X,Y\in L^\infty$ such that $$\p(X=x_0)=\p(X=-x_0)=\frac 14,~\p(X=y_0)=\frac 12,~\mbox{and}~Y=2X\id_{\{X=y_0\}}-X.$$
It follows that $$\VaR_{\Lambda_0}(X)=\VaR_{\Lambda_0}(Y)=\VaR_{\Lambda_0}\left(\frac{X+Y}{2}\right)=t_0.$$
For $x\in\R$, write
$$\begin{aligned}
   g(x)=cS_{\Lambda_0}(t_0,x)+f(x)&=c(x-t_0)_+-c\int^x_{t_0}(1-\Lambda_0(t))\,\d t+x\\
   &=\id_{\{x<0\}}\left((1-c(1-\alpha_3))x+c(1-\alpha_2)t_0\right)\\
   &+\id_{\{0\le x<t_0\}}\left((1-c(1-\alpha_2))x+c(1-\alpha_2)t_0\right)\\
   &+\id_{\{x\ge t_0\}}\left((1+c\alpha_1)x-c\alpha_1t_0\right),~~x\in\R.
\end{aligned}$$
Because $c\ge 1/(1-\sup_{x\in\R}\Lambda_0(x))$ by (ii), we have $c\ge 1/(1-\alpha_3)$ and thus $g(-x_0)+g(x_0)<2g(0)$. It follows that
$$\rho_{\Lambda_0}(X)=\rho_{\Lambda_0}(Y)=\frac{g(-x_0)+g(x_0)}{4}+\frac{g(y_0)}{2}<\frac{g(0)+g(y_0)}{2}=\rho_{\Lambda_0}\left(\frac{X+Y}{2}\right).$$
This leads to a contradiction to the quasi-convexity of $\rho_{\Lambda_0}$ and thus $\rho_\Lambda$ cannot be quasi-convex, normalized, and $\rho_\Lambda\ge \VaR_\Lambda$ simultaneously for all decreasing functions $\Lambda:\R\to(0,1)$.
\end{enumerate}

\end{document}